\documentclass[prd,twocolumn,twoside,preprintnumbers,superscriptaddress,nofootinbib]{revtex4-2}

\usepackage{physics}
\usepackage{amsmath}
\usepackage{amssymb}
\usepackage{hyperref}
\usepackage{enumitem}
\usepackage{xspace}
\usepackage{slashed}
\usepackage{braket}
\usepackage{leftindex}
\usepackage{graphicx}  %
\usepackage{bm}  %
\usepackage{amsmath}   %
\usepackage{graphics}
\usepackage{color}
\usepackage{colortbl}
\usepackage{xcolor}
\usepackage{comment}
\usepackage{titlesec}
\usepackage{enumitem}
\usepackage{setspace}
\usepackage{lineno}
\usepackage[normalem]{ulem}
\usepackage{soul}

\usepackage{graphicx}
\usepackage{array}
\usepackage{tabularray}
\usepackage[caption=false]{subfig}

\hypersetup{
    colorlinks=true,       
    linkcolor=blue,          
    citecolor=blue,        
    filecolor=blue,      
    urlcolor=blue           
}

\allowdisplaybreaks[1]

\titlespacing{\subsubsection}{0pt}{\parskip}{-\parskip}

\titlespacing{\section}{0pt}{2.5ex plus 1ex minus .2ex}{1.8ex plus .2ex}
\titlespacing{\subsection}{0pt}{2.25ex plus 1ex minus .2ex}{1ex plus .2ex}

\begin{document}

\title{Muon $g$$-$$2$: correlation-induced uncertainties in precision data combinations}

\author{Alexander~Keshavarzi}
\affiliation{Department of Physics and Astronomy, University College London, London WC1E 6BT, U.K.}

\author{Daisuke~Nomura}
\affiliation{Department of Radiological Sciences, International University of Health and Welfare, Tochigi 324-8501, Japan}

\author{Thomas~Teubner}
\affiliation{Department of Mathematical Sciences, University of Liverpool, Liverpool L69 3BX, U.K.}

\author{Aidan~Wright}
\affiliation{Department of Mathematical Sciences, University of Liverpool, Liverpool L69 3BX, U.K.}

\begin{abstract}
We present a general and systematic framework to quantify uncertainties arising from imperfectly known systematic correlations in data combinations. Formulated at the level of the combined data, the method enables controlled variation of the correlation structure, leading to the construction of covariance matrices directly on the resulting combination and thus providing a robust and systematic estimate of correlation-induced uncertainties. We apply the method to $e^+e^- \rightarrow \mathrm{hadrons}$ cross section data, with the resulting covariance matrices propagated to derived observables, including dispersive determinations of the hadronic vacuum polarization (HVP) contribution to the muon anomalous magnetic moment, $a_\mu^\mathrm{HVP}$. We find that uncertainties from systematic correlation assumptions are generally subdominant but non-negligible, and do not fully account for differences between existing $e^+e^- \rightarrow \mathrm{hadrons}$ data combinations. The framework is broadly applicable to correlated data combinations in precision measurements and constitutes a new component of the upcoming KNTW data combination for $a_\mu^\mathrm{HVP}$.
\end{abstract}

\maketitle
\section{INTRODUCTION}

Combining correlated measurements is a recurring and nontrivial scientific challenge. Generally, the correlation structure encodes essential information about shared statistical or systematic effects, common sources of uncertainty, or intrinsic dependencies between observables, and must be explicitly incorporated into any statistically rigorous combination procedure. While such approaches are well established, their reliability depends critically on how accurately the correlations -- and therefore the underlying variances -- are known. In practice, however, systematic uncertainties and their correlations are not directly measurable quantities, but rather estimates of the potential size and behavior of underlying effects. As a result, the associated correlation structure is intrinsically uncertain and often relies on assumptions or modeling choices. This inherent ambiguity means that different, equally plausible treatments of systematic correlations can lead to different combined results, introducing an additional source of systematic uncertainty that is not always explicitly quantified.

The impact of this effect is intrinsically linked to both the choice of combination methodology and the analysis decisions made in its implementation. Common approaches include likelihood-based fits such as $\chi^2$ minimization and Monte Carlo sampling techniques which may yield different weightings of correlated uncertainties and, therefore, different combination outcomes. Furthermore, analysts may impose constraints on how correlations are treated -- for example, limiting their magnitude, applying them globally across the full parameter space, or restricting them to local regions. Such choices, while often motivated by stability or interpretability, themselves constitute sources of systematic uncertainty. 

The treatment of correlations, particularly when their magnitude and structure are imperfectly known, may induce biases in the resulting combination. These effects can be particularly pronounced when combining measurements that are in tension, where correlations -- whether between distinct datasets or within a single measurement -- can shift weights and alter central values in a nontrivial and sometimes unintuitive manner. A tension-dominated environment exposes limitations in existing data combination procedures that incorporate correlations between measurements, and may require the introduction of additional conservative uncertainties to account for inconsistencies. 

This work addresses these issues in the context of determining the low-energy, non-perturbative hadronic cross section, a quantity that relies heavily on the combination of correlated data. The $e^+e^- \rightarrow \mathrm{hadrons}$ cross section ($\sigma_{\rm had}$) in this regime is typically obtained from measurements of exclusive hadronic final states for $\sqrt{s} \lesssim 2\,\mathrm{GeV}$, and from inclusive measurements of $e^+e^- \rightarrow \text{all hadrons}$ for $\sqrt{s} \gtrsim 2\,\mathrm{GeV}$, with each final state referred to as a “channel”. Within each channel, multiple measurements are generally combined using their covariance matrices to produce a single, best estimate of the cross section which, in turn, plays a critical role in precision calculations of hadronic contributions to observables such as the anomalous magnetic moment of the muon, $a_\mu$ (muon $g$$-$$2$). Dispersive approaches to determine the hadronic vacuum polarization (HVP) contributions, $a_\mu^\mathrm{HVP}$~\cite{Brodsky:1967sr,Lautrup:1968tdb,Krause:1996rf,Jegerlehner:2017gek,Jegerlehner:2015stw,Jegerlehner:2017lbd,Jegerlehner:2017zsb,Jegerlehner:2018gjd,Eidelman:1995ny,Benayoun:2007cu,Benayoun:2012etq,Benayoun:2012wc,Benayoun:2015gxa,Benayoun:2019zwh,Benayoun:2019zwh,Davier:2010nc,Davier:2010nc,Hagiwara:2003da,Hagiwara:2006jt,Hagiwara:2011af,davier:2017zfy,kurz:2014wya,davier:2019can,kurz:2014wya,keshavarzi:2018mgv,colangelo:2018mtw,hoferichter:2019mqg,keshavarzi:2019abf,melnikov:2003xd,masjuan:2017tvw,Colangelo:2017fiz,hoferichter:2018kwz,keshavarzi:2024bli}, rely on these combinations of correlated $\sigma_{\rm had}$ measurements as input. However, significant tensions between these inputs to evaluations of $a_\mu^\mathrm{HVP}$ limit the ability to achieve a Standard Model (SM) prediction, $a_\mu^\mathrm{SM}$~\cite{Aliberti:2025qei,davier:2017zfy,keshavarzi:2018mgv,colangelo:2018mtw,hoferichter:2019mqg,davier:2019can,keshavarzi:2019abf,kurz:2014wya,melnikov:2003xd,masjuan:2017tvw,Colangelo:2017fiz,hoferichter:2018kwz,DiLuzio:2024sps,keshavarzi:2024bli,blum:2018mom, Giusti:2019xct, Borsanyi:2020mff, Lehner:2020crt, Wang:2022lkq, Aubin:2019usy, Ce:2022kxy, ExtendedTwistedMass:2022jpw,RBC:2023pvn,Kuberski:2024bcj,Boccaletti:2024guq,Spiegel:2024dec,RBC:2024fic,Djukanovic:2024cmq,ExtendedTwistedMass:2024nyi,MILC:2024ryz,FermilabLatticeHPQCD:2024ppc,RBC:2018dos,Aubin:2022hgm,Ludtke:2024ase,colangelo:2014qya,aoyama:2012wk,czarnecki:2002nt,gnendiger:2013pva,bijnens:2019ghy,Blum:2019ugy,Volkov:2019phy,Volkov:2024yzc,Aoyama:2024aly,Parker:2018vye,Morel:2020dww,Fan:2022eto,Hoferichter:2025yih,Colangelo:2015ama,Eichmann:2019tjk,Leutgeb:2019gbz,Cappiello:2019hwh,Masjuan:2020jsf,Bijnens:2020xnl,Bijnens:2021jqo,Danilkin:2021icn,Stamen:2022uqh,Leutgeb:2022lqw,Hoferichter:2023tgp,Hoferichter:2024fsj,Estrada:2024cfy,Deineka:2024mzt,Eichmann:2024glq,Bijnens:2024jgh,Hoferichter:2024bae,Holz:2024diw,Cappiello:2025fyf,Chao:2021tvp,Chao:2022xzg,Blum:2023vlm,Fodor:2024jyn}, that has a precision commensurate with the final, 127 parts-per-billion (ppb) measurement of $a_\mu^{\rm exp}$~\cite{Muong-2:2025xyk} by the Fermilab Muon $g$$-$$2$ experiment~\cite{Muong-2:2002wip, Muong-2:2004fok, Muong-2:2006rrc, Muong-2:2021ojo, Muong-2:2021ovs, Muong-2:2021vma, Muong-2:2021xzz, Muong-2:2023cdq, Muong-2:2024hpx}.

\begin{figure}[!t]
    \centering
    \includegraphics[width=\linewidth]{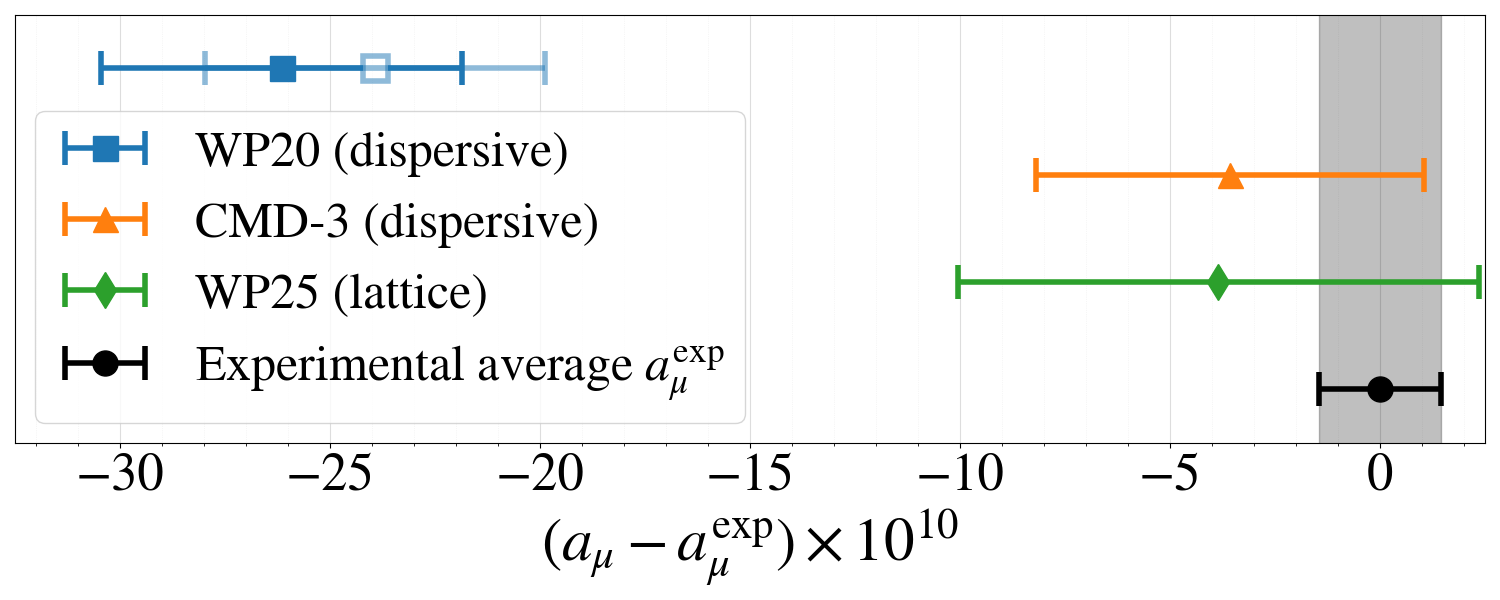}
    \caption{A comparison of different values of $a_\mu$, shown as differences with the measured value $a_\mu^{\rm exp}$~\cite{Muong-2:2025xyk,Muong-2:2002wip, Muong-2:2004fok, Muong-2:2006rrc, Muong-2:2021ojo, Muong-2:2021ovs, Muong-2:2021vma, Muong-2:2021xzz, Muong-2:2023cdq, Muong-2:2024hpx}, indicated by the black circle and gray uncertainty band. The other markers correspond to Standard Model predictions $a_\mu^{\rm SM}$ based on different determinations of $a_\mu^\mathrm{HVP}$~\cite{Muong-2:2002wip, Muong-2:2004fok,Muong-2:2006rrc,Muong-2:2021ojo,Muong-2:2021ovs,Muong-2:2021vma,Muong-2:2021xzz,Muong-2:2023cdq,Muong-2:2024hpx,Muong-2:2025xyk,Aliberti:2025qei,davier:2017zfy,keshavarzi:2018mgv,colangelo:2018mtw,hoferichter:2019mqg,davier:2019can,keshavarzi:2019abf,kurz:2014wya,melnikov:2003xd,masjuan:2017tvw,Colangelo:2017fiz,hoferichter:2018kwz,DiLuzio:2024sps,colangelo:2014qya,keshavarzi:2024bli,blum:2018mom, Giusti:2019xct, Borsanyi:2020mff, Lehner:2020crt, Wang:2022lkq, Aubin:2019usy, Ce:2022kxy, ExtendedTwistedMass:2022jpw,RBC:2023pvn,Kuberski:2024bcj,Boccaletti:2024guq,Spiegel:2024dec,RBC:2024fic,Djukanovic:2024cmq,ExtendedTwistedMass:2024nyi,MILC:2024ryz,FermilabLatticeHPQCD:2024ppc,RBC:2018dos,Aubin:2022hgm,Ludtke:2024ase,aoyama:2012wk,czarnecki:2002nt,gnendiger:2013pva,bijnens:2019ghy,Blum:2019ugy,Volkov:2019phy,Volkov:2024yzc,Aoyama:2024aly,Parker:2018vye,Morel:2020dww,Fan:2022eto,Hoferichter:2025yih,Colangelo:2015ama,Eichmann:2019tjk,Leutgeb:2019gbz,Cappiello:2019hwh,Masjuan:2020jsf,Bijnens:2020xnl,Bijnens:2021jqo,Danilkin:2021icn,Stamen:2022uqh,Leutgeb:2022lqw,Hoferichter:2023tgp,Hoferichter:2024fsj,Estrada:2024cfy,Deineka:2024mzt,Eichmann:2024glq,Bijnens:2024jgh,Hoferichter:2024bae,Holz:2024diw,Cappiello:2025fyf,Chao:2021tvp,Chao:2022xzg,Blum:2023vlm,Fodor:2024jyn,Aoyama:2020ynm,aoyama:2012wk, Aoyama:2019ryr,czarnecki:2002nt,gnendiger:2013pva,davier:2017zfy,keshavarzi:2018mgv,colangelo:2018mtw,hoferichter:2019mqg,davier:2019can,keshavarzi:2019abf,kurz:2014wya,melnikov:2003xd,masjuan:2017tvw,Colangelo:2017fiz,hoferichter:2018kwz,gerardin:2019vio,bijnens:2019ghy,colangelo:2019uex,Blum:2019ugy,CMD-3:2023alj,CMD-3:2023rfe}. The green diamond denotes the Muon $g$$-$$2$ Theory Initiative (TI)~\cite{Aliberti:2025qei,Aoyama:2020ynm} 2025 White Paper (WP25)~\cite{Aliberti:2025qei} result, based on lattice QCD determinations of $a_\mu^\mathrm{HVP}$~\cite{RBC:2018dos, Giusti:2019xct,Borsanyi:2020mff,Lehner:2020crt,Wang:2022lkq,Aubin:2022hgm,Ce:2022kxy,ExtendedTwistedMass:2022jpw,RBC:2023pvn,Kuberski:2024bcj,Boccaletti:2024guq,Spiegel:2024dec,RBC:2024fic,Djukanovic:2024cmq,ExtendedTwistedMass:2024nyi,MILC:2024ryz,FermilabLatticeHPQCD:2024ppc,Aubin:2019usy}. The orange triangle shows a dispersive evaluation using CMD-3 $\sigma_{\pi\pi}$ data~\cite{CMD-3:2023alj, CMD-3:2023rfe} in isolation, supplemented by other channels and energy regions as described in~\cite{Aliberti:2025qei}. The blue squares correspond to the TI's 2020 White Paper result (WP20)~\cite{Aoyama:2020ynm,aoyama:2012wk, Aoyama:2019ryr,czarnecki:2002nt,gnendiger:2013pva,davier:2017zfy,keshavarzi:2018mgv,colangelo:2018mtw,hoferichter:2019mqg,davier:2019can,keshavarzi:2019abf,kurz:2014wya,melnikov:2003xd,masjuan:2017tvw,Colangelo:2017fiz,hoferichter:2018kwz,gerardin:2019vio,bijnens:2019ghy,colangelo:2019uex,Blum:2019ugy,colangelo:2014qya}, based solely on dispersive determinations of $a_\mu^\mathrm{HVP}$~\cite{davier:2017zfy,keshavarzi:2018mgv,colangelo:2018mtw,hoferichter:2019mqg,davier:2019can,keshavarzi:2019abf,kurz:2014wya}. The solid square corresponds exactly to WP20~\cite{Aoyama:2020ynm}, while the open square uses WP20 HVP contributions combined with updated non-HVP SM contributions from WP25~\cite{Aliberti:2025qei}.}
    \label{fig:amu_comparison}
    \vspace{-0.4cm}
\end{figure}
A discrepancy exists between the dispersive evaluations of $a_\mu^\mathrm{HVP}$ and those obtained from lattice QCD~\cite{RBC:2018dos,Giusti:2019xct,Borsanyi:2020mff,Lehner:2020crt,Wang:2022lkq,Aubin:2022hgm,Ce:2022kxy,ExtendedTwistedMass:2022jpw,RBC:2023pvn,Kuberski:2024bcj,Boccaletti:2024guq,Spiegel:2024dec,RBC:2024fic,Djukanovic:2024cmq,ExtendedTwistedMass:2024nyi,MILC:2024ryz,FermilabLatticeHPQCD:2024ppc,Aubin:2019usy,Beltran:2026ofp}, with the latter generally yielding higher values (see Fig.~\ref{fig:amu_comparison}). Within the dispersive framework itself, additional tensions arise from inconsistencies between input datasets, which in some cases are further compounded by differing analysis choices made in the compilation of $\sigma_{\rm had}$ by different groups. This is most notable in the dominant $e^+e^- \rightarrow \pi^+\pi^-$ channel, the cross section $\sigma_{\pi\pi}$ of which contributes $\sim75\%$ of the leading order $a_\mu^\mathrm{HVP}$ and therefore has a substantial influence on the SM prediction. 

\begin{figure}[!t]
    \centering
    \includegraphics[width=\linewidth]{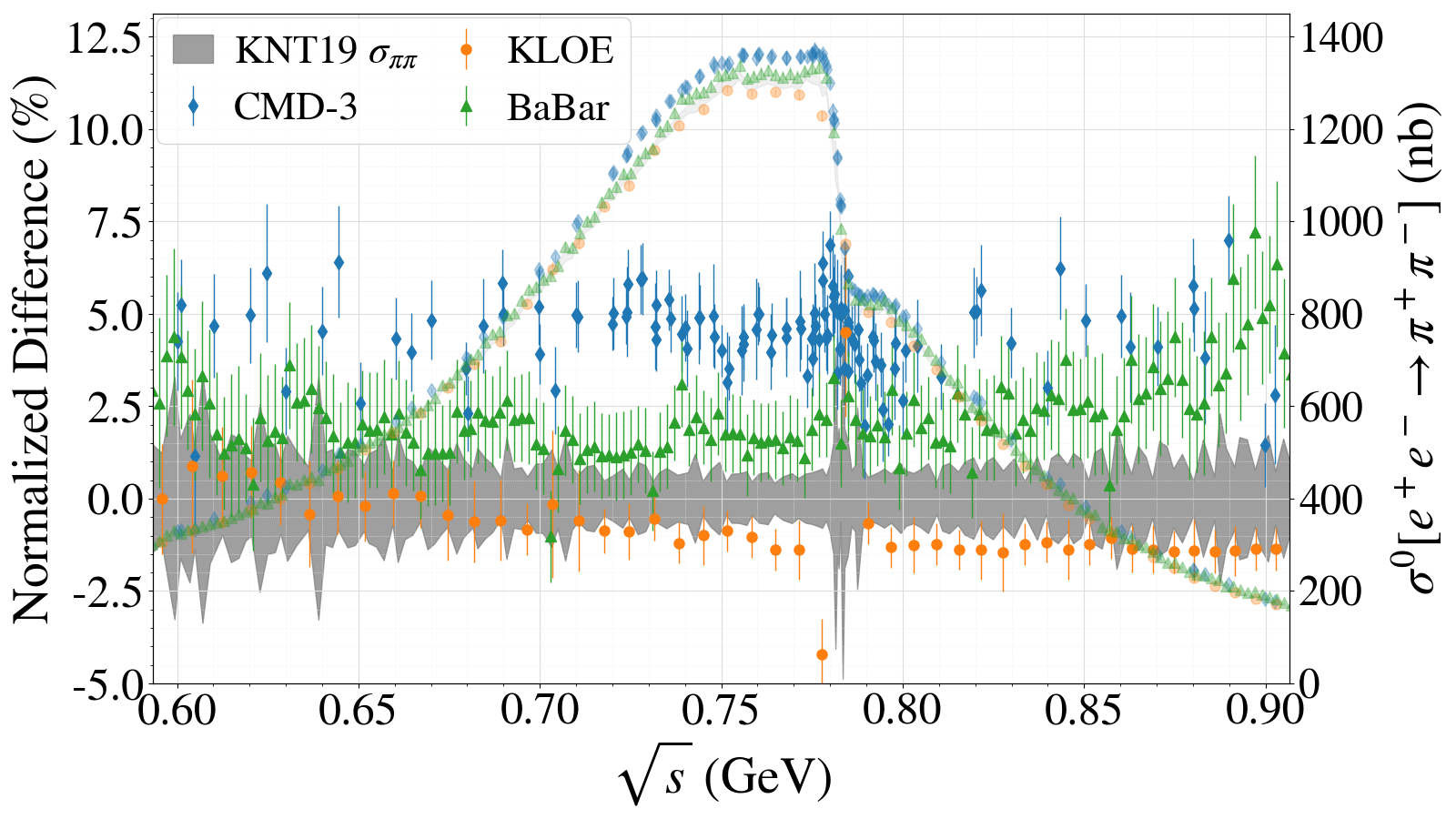}
    \caption{Comparison of the results of the CMD-3~\cite{CMD-3:2023alj, CMD-3:2023rfe}, KLOE~\cite{KLOE:2008fmq, KLOE:2010qei, KLOE:2012anl, KLOE-2:2017fda} and BaBar~\cite{BaBar:2009wpw,BaBar:2012bdw} measurements of the $\sigma_{\pi\pi}$ spectrum to the combined KNT19 $\sigma_{\pi\pi}$ values~\cite{keshavarzi:2019abf} which did not include CMD-3. The differences normalized to KNT19 are shown on the left axis whereas the (faded) unnormalized cross sections are shown on the right axis.}
    \label{fig:spectra_comparison}
    \vspace{-0.4cm}
\end{figure}

Historically, while such measurements have achieved high precision~\cite{akhmetshin:2003zn, aulchenko:2006na, akhmetshin:2006wh, akhmetshin:2006bx, achasov:2006vp, BaBar:2009wpw,BaBar:2012bdw, KLOE:2008fmq, KLOE:2010qei, KLOE:2012anl, KLOE-2:2017fda, BESIII:2015equ,CMD-3:2023alj, CMD-3:2023rfe}, persistent tensions between input data already posed challenges -- preeminently the long-standing discrepancy between the BaBar~\cite{BaBar:2009wpw,BaBar:2012bdw} and KLOE~\cite{KLOE:2008fmq, KLOE:2010qei, KLOE:2012anl, KLOE-2:2017fda} measurements of $\sigma_{\pi\pi}$. More recently, the situation has been significantly exacerbated by a newer CMD-3 measurement~\cite{CMD-3:2023alj, CMD-3:2023rfe}, which, as shown in Fig.~\ref{fig:spectra_comparison}, lies systematically higher than the existing body of $\pi^+\pi^-$ data. Even since then, the disagreement between increasingly precise $\sigma_{\pi\pi}$ measurements continues to worsen, with recent announcements of preliminary results further highlighting the tension: the latest SND measurement is set to exhibit a significant upward shift appearing to corroborate the higher values observed by CMD-3~\cite{SND2025}, while a new BaBar measurement will show remarkable agreement with its predecessor~\cite{BaBar2025}.

These discrepancies have profound implications. As shown in Fig.~\ref{fig:amu_comparison}, the CMD-3 spectrum, when used in isolation as input to dispersive evaluations, yields values of $a_\mu^\mathrm{SM}$ that are consistent with lattice QCD determinations and in agreement with the experimental measurement. In contrast, combinations based on the previous $\sigma_{\pi\pi}$ data lead to significantly lower values of $a_\mu^\mathrm{SM}$, maintaining the long-standing tension with experiment. As a result, current evaluations of $a_\mu^\mathrm{SM}$ span a range that covers both agreement with $a_\mu^\mathrm{exp}$ and significant deviation potentially indicative of new physics, highlighting the sensitivity of the result to the treatment and combination of input data.

While ongoing experimental and theoretical efforts -- including new $\sigma_{\rm had}$ measurements~\cite{BaBar2025,BelleII2025, BESIII2025,CMD32025,KLOE2025,SND2025}, improved radiative corrections~\cite{Strong2020,Aliberti:2024fpq,PetitRosas:2026iuq,Budassi:2024whw,Budassi:2026lmr,CarloniCalame:2026hhy}, and complementary approaches such as the MUonE experiment~\cite{CarloniCalame:2015obs,Abbiendi:2016xup,Abbiendi:2677471} -- aim to clarify the underlying physics, it is important to understand the role of data combination methodologies in shaping current and future results. In particular, the interplay between correlation modeling, methodological choices, and dataset inconsistencies introduces an additional layer of systematic uncertainty that should be quantified.

In this paper, we investigate the impact of these effects on combinations of $\sigma_{\rm had}$ measurements, explicitly quantifying how changes in the correlation structure, together with discrepancies in the input data, propagate to the resulting cross section. We further explore the dependence of the result on methodological choices and assumptions about systematic uncertainties. The study is performed at the level of the combined cross section itself, rather than on derived or integrated observables, thereby retaining local variations and tensions in the data and enabling their consistent propagation regardless of the application. 

More generally, we introduce a systematic and flexible framework to quantify uncertainties arising from imperfectly known systematic correlations in data combinations. Within this framework, additional systematic uncertainties introduced by the combination procedure are determined and propagated to derived observables such as $a_\mu^\mathrm{HVP}$, providing a more robust and transparent tool for precision analyses. Although the methodology is developed and demonstrated in the context of low energy $\sigma_\text{had}$ measurements, it is directly applicable to a broad class of problems involving correlated datasets.

This paper is organized as follows. Sec.~\ref{sec:corr_comb} provides an overview of correlations in $\sigma_\mathrm{had}$ data and, in particular, their use in combinations to determine derived observables such as $a_\mu^\mathrm{HVP}$. In Sec.~\ref{sec:decorr_method}, we introduce a general procedure to evaluate uncertainties resulting from assumptions about systematic correlations. Sec.~\ref{application} quantifies the impact of this new procedure on the KNT19 combinations of $\sigma_\mathrm{had}$ data and corresponding derived observables~\cite{keshavarzi:2019abf}. Following the conclusions in Sec.~\ref{sec:conclusions}, and in light of these findings, we provide in Appendix~\ref{Appendix} an explanation for a historic discrepancy between the DHMZ19~\cite{davier:2019can} and KNT19~\cite{keshavarzi:2019abf} combinations.

\section{CORRELATIONS IN $\sigma_\text{had}$ DATA COMBINATIONS} \label{sec:corr_comb}

As described above, dispersive calculations of $a_\mu^\text{HVP}$ rely on channel-by-channel combinations of measured $\sigma_{\rm had}$ data. These measurements are usually reported as binned cross sections with statistical and systematic uncertainties provided separately. Where correlations are present, experiments may supply corresponding covariance matrices; alternatively, these must be constructed based on knowledge or assumptions about the correlation structure of the channel.\footnote{Note that correlations can also exist between measurements in different channels, although there are no instances of experiments providing such covariance matrices. However, whilst estimates of these correlations have been used to quantify an additional uncertainty on derived quantities~\cite{Davier:2010nc,Davier:2010rnx,davier:2017zfy,davier:2019can,Davier:2023fpl}, their full use in data combinations through appropriately constructed covariance matrices would require a global fit of all channels.} Consequently, measurements that feature in a combination have one or more covariance matrices $C_{ij}=\rho_{ij}\,d\sigma_id\sigma_j$, where $d\sigma_{i/j}$ is an uncertainty on $i^\text{th}$/$j^\text{th}$ cross section value and $-1 \leq \rho_{ij} \leq 1$ is the correlation coefficient. The case $\rho_{ij}=1$ is referred to in this work as full (100\%) correlation and the converse $\rho_{ij}=\delta_{ij}$ is referred to as full decorrelation, where $\delta_{ij}$ is the Kronecker delta. Where experimental uncertainties on measurements are divided into statistical (stat) and systematic (sys), the total (tot) covariance matrix for a given measurement is defined as
\begin{align} \label{eq:Ctot}
    C_{\rm tot} & = C_{\rm stat} + C_{\rm sys} \\ \nonumber
    & = \rho^{\rm stat}_{ij}d\sigma^{\rm stat}_i d\sigma^{\rm stat}_j + \sum_{m}\rho^{\text{sys},m}_{ij}d\sigma^{\text{sys},m}_i d\sigma^{\text{sys},m}_j \, ,
\end{align}
where the sum runs over different sources of systematic uncertainty, $m$.

The statistical correlation coefficient $\rho_{ij}^\mathrm{stat}$ is determined directly from the underlying statistics of a measurement. In $e^+e^- \to \mathrm{hadrons}$ measurements, statistical uncertainties are typically uncorrelated. However, procedures such as unfolding~\cite{DAgostini:1994fjx, Hocker:1995kb, Malaescu:2009dm, DAgostini:2010hil}, which correct measured cross sections for detector effects (e.g. resolution, efficiency, and smearing), or the use of common luminosity normalization across multiple bins, can introduce non-zero -- and in some cases large -- statistical correlations between cross section values. An example of such effects is discussed in Appendix~\ref{Appendix}. Importantly, these correlations arise from well-defined statistical processes, such as the redistribution of events, and therefore do not rely on external assumptions. As such, they can be treated as robust and reliable inputs in data combination procedures.

Conversely, systematic uncertainties are not derived quantities and are instead estimates of the size of effects, meaning the level to which they and their correlations are known is limited. Some sources of systematic correlations arise from normalization uncertainties. In the context of $e^+e^-\to\text{hadrons}$, these can be uncertainties e.g. on efficiencies, luminosity and radiative corrections. These apply globally, as an erroneous shift at one point due to such a systematic effect implies a shift at all others, motivating an assumption $\rho_{ij}^{\text{sys},m}=1$. Energy dependence in systematic correlations can be incorporated if such information is provided. Following Eq.~\eqref{eq:Ctot}, covariance matrices for a given measurement are a sum of all different contributions. 

Even when individual systematic uncertainties are treated as fully correlated, this does not imply that all regions of a spectrum are equally correlated. In practice, different systematic effects dominate in different energy regions, and their relative sizes can vary significantly. As a result, the total uncertainty in each region is composed differently, leading to a reduced effective correlation between regions. Consequently, parts of a spectrum which are subject to distinct or varying levels of systematic effects are naturally less correlated. In extreme cases, such as resonant channels like $e^+e^- \rightarrow \pi^+\pi^-\pi^0$, regions with different physical or analysis characteristics can be effectively uncorrelated.

The KNTW~\cite{keshavarzi:2024bli} (previously KNT~\cite{keshavarzi:2018mgv,keshavarzi:2019abf}) compilations of the total hadronic cross section provide one of the three dispersive inputs to $a_\mu^\mathrm{HVP}$ used by the TI to determine $a_\mu^{\rm SM}$~\cite{Aliberti:2025qei,Aoyama:2020ynm}, alongside CHKLS~\cite{colangelo:2018mtw,hoferichter:2019mqg,Colangelo:2022prz,Stoffer:2023gba,Hoferichter:2025lcz,Leplumey:2025kvv} and DHMZ~\cite{Davier:2010nc, Davier:2010rnx, davier:2017zfy, davier:2019can, Davier:2023fpl}. As discussed in~\cite{keshavarzi:2024bli}, differing analysis choices and assumptions in these combination procedures -- particularly in the treatment of correlations in data fits -- can lead to variations in the resulting cross sections and, consequently, in $a_\mu^\mathrm{HVP}$ across individual channels. To account for this, WP20~\cite{Aoyama:2020ynm} adopted a conservative averaging procedure, introducing additional case-by-case systematic uncertainties to reflect tensions both between experimental spectra and between the resulting dispersive evaluations of $a_\mu^\mathrm{HVP}$ (which are not always independent sources of tension). Importantly, these additional uncertainties were not evaluated at the level of the combined cross sections, but rather inferred from differences in the integrated values of $a_\mu^\mathrm{HVP}$, thereby reducing sensitivity to local variations and tensions present in cross sections themselves. The dominant contribution to the WP20 error budget on $a_\mu^\mathrm{SM}$ arose from an additional systematic uncertainty -- the “BaBar–KLOE” uncertainty -- introduced to account for the discrepancy between the BaBar~\cite{BaBar:2009wpw,BaBar:2012bdw} and KLOE~\cite{KLOE:2008fmq, KLOE:2010qei, KLOE:2012anl, KLOE-2:2017fda} determinations of $a_\mu^{\pi^+\pi^-}$ derived from their $\sigma_{\pi\pi}$ measurements, as well as the resulting differences between the CHKLS, DHMZ19 and KNT19 evaluations of $a_\mu^{\pi^+\pi^-}$~\cite{Aoyama:2020ynm}.

In the KNTW data combination procedure, all experimental correlation information is maximally used to influence the mean values (whilst carefully avoiding the d'Agostini bias~\cite{DAgostini:1993arp}). This approach assumes that all data provided by an experiment constitute, as accurately as possible, information that reflects the measurement conditions and data analysis. Where explicit information about the systematic correlation structure is not provided by the experimental collaborations, the assumption $\rho_{ij}^{\text{sys,}m} = 1$ is consistently applied. The CHKLS group~\cite{colangelo:2018mtw, hoferichter:2019mqg,Hoferichter:2025lcz,Leplumey:2025kvv} employ similar correlation assumptions. In contrast, the DHMZ data combination~\cite{davier:2019can, Davier:2010rnx} propagates covariances to the final uncertainties, but restricts their influence on the mean values to local regions, thereby avoiding the implicit assumption that the provided covariances are perfectly known~\cite{Aoyama:2020ynm}. This allows data points to fit locally without being constrained by information from non-local regions, and provides a clear example of how differing analysis choices can lead to different results. 

Historically, this difference in the treatment of correlations was understood to be a significant source of discrepancies between the KNTW (CHKLS) and DHMZ data combinations, resulting cross sections, and values of $a_\mu^\mathrm{HVP}$. Further, it was believed to have driven many of the differences observed in WP20 across individual channels. In particular, it has been assumed that these differing treatments of systematic correlations effectively map the underlying BaBar–KLOE tension in $\sigma_{\pi\pi}$ data onto the combined results, leading to DHMZ–KNTW differences in $a_\mu^{\pi^+\pi^-}$ that mirror the spread between the BaBar and KLOE measurements themselves, and subsequently entered WP20~\cite{keshavarzi:2019abf,davier:2019can,Aoyama:2020ynm}. 

In this work, we develop a generalized procedure, formulated at the level of the data combination, to quantify the impact of combining data under varying correlation assumptions and to systematically evaluate a resulting uncertainty. Within this framework, the effects of differing correlation treatments -- such as those underlying the KNTW and DHMZ approaches -- are quantified for the first time in a non-ad hoc manner.

\section{CORRELATION VARIATION METHODOLOGY}\label{sec:decorr_method}

In this section, we present a general procedure to evaluate uncertainties on a spectrum resulting from assumptions about systematic correlations. First a \textit{measure of deviation}, which quantifies a change in a spectrum when varying the underlying correlation assumptions, is constructed and discussed. The means to estimate a systematic uncertainty on a spectrum using this measure are also introduced. Then, different \textit{decorrelation procedures} are discussed; these allow the initial correlation assumptions to be varied for the purposes of the measure and uncertainty evaluation. A simplified example is used to demonstrate estimation of a \textit{correlation-strength systematic uncertainty}, $d^\rho$, using the measure and decorrelation procedures. 

\subsection{The measure of deviation}

Consider a general $n$-dimensional spectrum $O_{\underline{i}}$, which is a combination of input datasets. Here $\underline{i}$ labels the indices of $O$; for example a single index could run over the energy bins $\sqrt{s}_i$ of a cross section combination $\sigma_i$. A multiplet of indices could represent the labels and parameters of parton distributions. Any data combination necessarily makes assumptions about the (systematic) correlations between data points, encoded in a covariance matrix $C_{\underline{i}\underline{j}}$. Different assumptions $\tilde C_{\underline{i}\underline{j}}$ typically yield different combinations $\tilde O_{\underline{i}}$, motivating the computation of an associated systematic uncertainty. If the default combination is $O_{\underline{i}}$ and the most different combination is $O'_{\underline{i}}$, a new systematic covariance matrix $C^\rho_{\underline{i} \underline{j}}$ can be defined as
\begin{equation} \label{eq:Crho}
    C^\rho_{\underline{i}\underline{j}}=\Bigl[O_{\underline{i}}'-O_{\underline{i}}\Bigr] \Bigl[O_{\underline{j}}'-O_{\underline{j}}\Bigr].
\end{equation}

To obtain $O_{\underline{i}}'$, we construct a measure $\mathcal{M}$ such that $\mathcal{M}(O_{\underline{i}}')=\max[\mathcal{M}(\tilde O_{\underline{i}})]$, i.e.\ $\mathcal{M}$ is maximized when the deviation of $\tilde O_{\underline{i}}$ from $O_{\underline{i}}$ is maximized. The measure should have the following properties: (i) depend on the absolute values of deviations in the spectrum, (ii) be normalized so as to be dimensionless, (iii) be finite for all $\underline{i}$, and (iv) preferably be symmetric in $O_{\underline{i}}$ and $\tilde O_{\underline{i}}$. Our chosen measure, fulfilling all of these properties, is
\begin{equation}
    \mathcal{M} = \sum_{\underline{i},\underline{j}}\left[\left(\tilde{O}_{\underline{i}} -O_{\underline{i}}\right)\left(\tilde{C}_{\underline{i}\underline{j}} + {C}_{\underline{i}\underline{j}}\right)^{-1} \left(\tilde{O}_{\underline{j}} - O_{\underline{j}}\right)\right]. \label{eq:measure}
\end{equation}
Here the normalization is provided by the covariance matrices. If their derivation is unfeasible, or their inversion untenable, a simpler measure normalized by squared uncertainties could instead be employed. Alternatively, an even simpler measure based solely on the spectra is $\sum_{\underline{i}}[(\tilde{O}_{\underline{i}} - O_{\underline{i}})/(\tilde{O}_{\underline{i}} + O_{\underline{i}})]^2$. Studies have found only minor discrepancies between the results when using different measures.

All discussed choices of measure ensure a conservative uncertainty on the principal quantity: the resulting spectrum (e.g.\ of cross section values). However, for uncertainties on derived observables (such as $a_\mu^\text{HVP}$ obtained from a dispersion integral), positive and negative fluctuations may partially cancel under integration. It should also be noted that, since typical assumptions about the nature of systematic correlations do not necessarily yield full rank covariance matrices, inversion may be difficult. This can be avoided with standard techniques, or by including the statistical covariance matrices in the inverted $C_{\underline{i}\underline{j}}$ term. An inversion may also become increasingly untenable if required for high-dimensional $\underline{i}$.\footnote{The inverse of a rank $>2$ tensor is typically not well-defined. In such a case it is possible to `flatten' $\underline{i}$ to one dimension, invert the resulting matrix and then re-order. For $n>1$ dimensions, this will rapidly become computationally expensive.} Lastly, it is important to emphasize that Eq.~\eqref{eq:measure} explicitly depends on the choice of baseline, $O_{\underline{i}}$, i.e. a specific data combination procedure. Comparisons performed with respect to a different data combination would naturally yield different values of $\mathcal{M}$ for the same underlying input data. However, since the aim here is to determine a systematic uncertainty associated with the chosen baseline $O_{\underline{i}}$, this approach is internally consistent.

\subsection{Decorrelation procedures}

For most physical applications, including dispersive HVP studies, it is not practical  to scan the full parameter space of permissible modified covariance matrices $\tilde C_{\underline{i}\underline{j}}$. It is instead more attainable (and well-motivated) to consider how $C_{\underline{i}\underline{j}}$ may be modified in terms of physically motivated decorrelation scenarios. A general decorrelation procedure $F_{\underline{i}\underline{j}}(\boldsymbol{\alpha})$ can be defined as
\begin{equation}\label{eq:decorr_general}
    \tilde{C}_{\underline{i}\underline{j}}=F_{\underline{i}\underline{j}}(\boldsymbol{\alpha}) C_{\underline{i}\underline{j}} \, ,
\end{equation}
where $\boldsymbol{\alpha}=(\alpha_1,\,\alpha_2,\,...)$ are a set of \textit{decorrelation parameters}. These parameterize the collective behavior of the correlation coefficients. A scan over the space of each of these coefficients produces a space of $\tilde C_{\underline{i}\underline{j}}$ and thus $\tilde O_{\underline{i}}$ values where the maximum of $\mathcal{M}$ can be found and a systematic uncertainty estimated.

While a general decorrelation procedure can depend on multiple decorrelation parameters in a non-trivial way, here a factorized form
\begin{equation}\label{eq:decorr_factorised}
    \tilde{C}_{\underline{i}\underline{j}} = \biggl[\prod_{p}f^p_{\underline{i}\underline{j}}(\alpha^p)\biggr]C_{\underline{i}\underline{j}}
\end{equation}
is considered, in which different one-parameter decorrelation functions $f^p_{\underline{i}\underline{j}}(\alpha^p)$ are applied to the assumed covariance matrix. Such an approach is sufficient to describe many physical and well-motivated decorrelation scenarios. Particularly, it could be used to simulate the decorrelation of $C_{\underline{i}\underline{j}}$ independently along different dimensions of $\underline{i}$, to decorrelate statistics and systematics separately (see Appendix~\ref{Appendix}), or to perform different decorrelations for different systematic uncertainties.

\begin{figure}[!t]
    \centering
    \includegraphics[width=\linewidth]{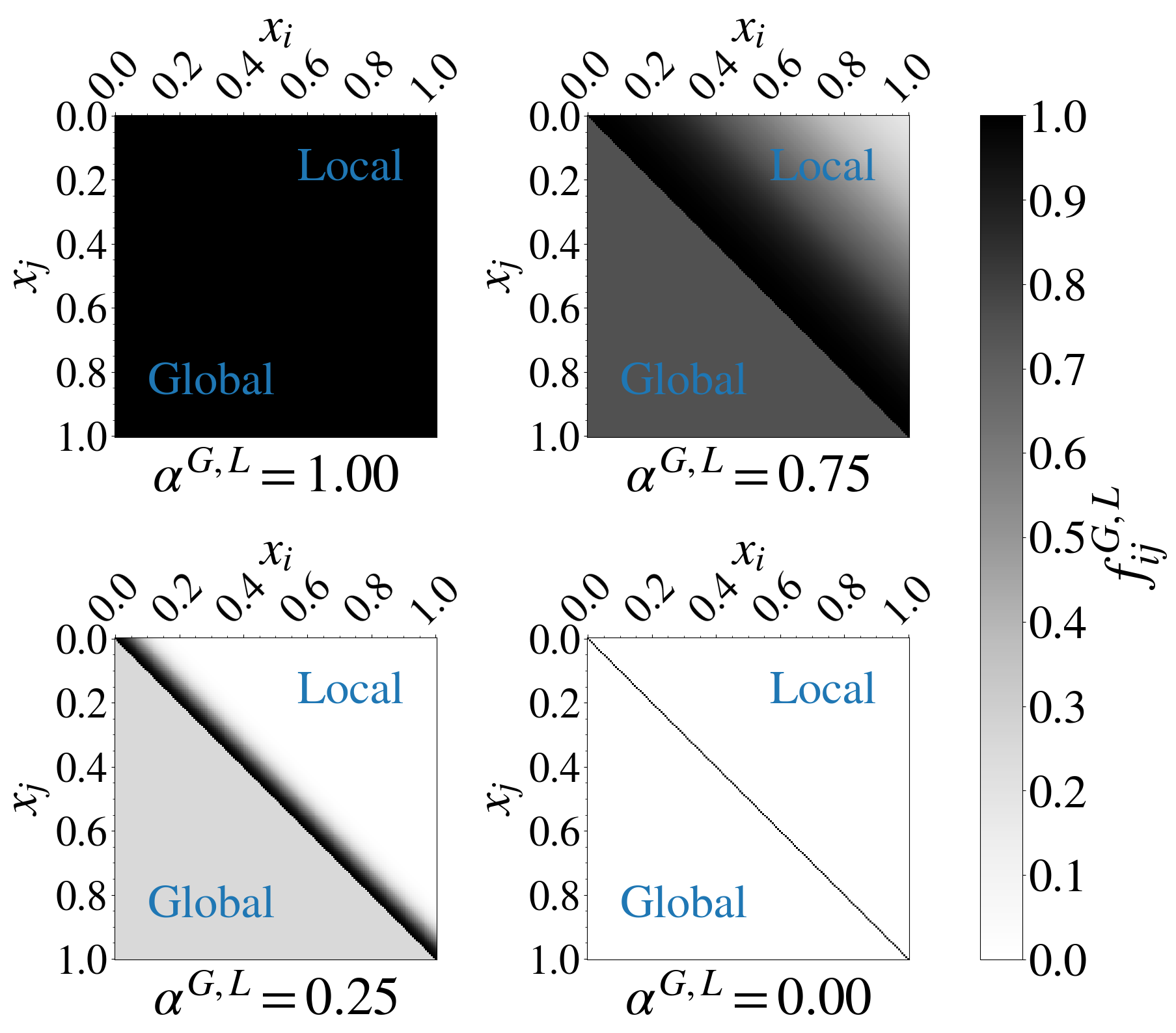}
    \caption{Global and local decorrelation functions $f_{ij}^\text{G,\,L}$ as defined in Eqs.~\eqref{eq:global_decorr} and~\eqref{eq:local_decorr} for arbitrary abscissa $x_i$. The values of $f_{ij}^\text{G,\,L}$ are diagonally symmetric and, for ease of comparison, are shown below and above the leading diagonal respectively. For local decorrelation, $\mathcal{E}=1/4$ is used.}
    \label{fig:local_decorr_visualisation}
    \vspace{-0.4cm}
\end{figure}

Specializing to the one-dimensional case and assuming $C_{ij}$ is taken with $\rho_{ij}^\text{sys,m}$ typically equal to 1 (though the generalizations should be straightforward), two example decorrelation procedures are presented: \textit{global} (G) and \textit{local} (L). In both cases, $\alpha=1$ is the default assumption and $\alpha=0$ is full decorrelation; both procedures are illustrated in Fig.~\ref{fig:local_decorr_visualisation}. The global decorrelation
\begin{equation}\label{eq:global_decorr}
    f^\text{G}_{ij}(\alpha^\text{G})=\alpha^\text{G}+(1-\alpha^\text{G})\,\delta_{ij} \, ,
\end{equation}
parameterized by $\alpha^\text{G}$,
is well-motivated by the above discussion of normalization-type correlations. All correlations are reduced by a constant factor $\alpha^\text{G}$ and the full range of possible factors is scanned.\footnote{Artificially imposing anticorrelations, while possible, is not explicitly considered in these decorrelation scenarios. Anticorrelations are instead accounted for where provided by experimental analyses.}
\begin{figure*}[!t] 
    \centering
    \includegraphics[width=0.75\textwidth]{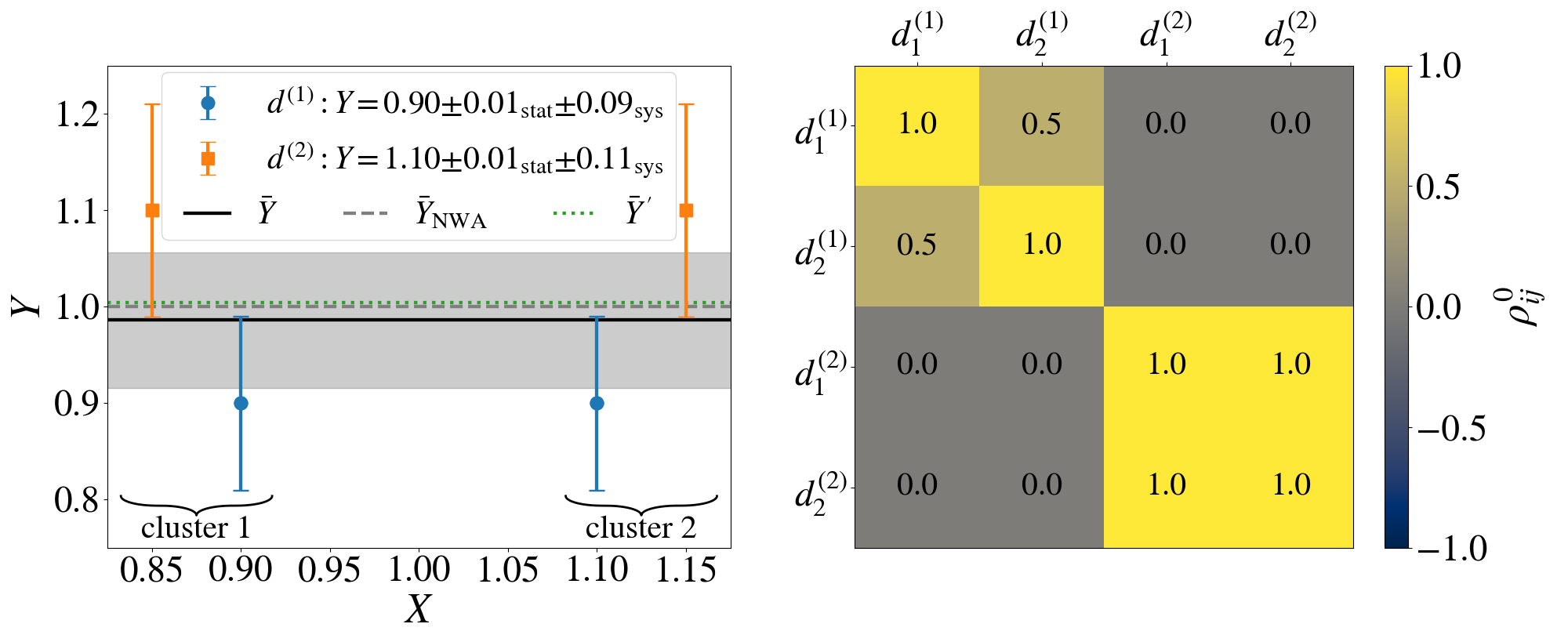}
    \caption{Toy model to demonstrate the measure and decorrelation procedures: two datasets $d^{(1,2)}$ each consisting of two points $d_{1,2}$ are to be combined into a single description. The common values $Y$ and uncertainties for each dataset are indicated in the left panel (also showing the data) and the systematic correlation coefficients between data points are shown in the right panel. The KNTW combination $\bar{Y}$ is shown as a solid black line with a shaded uncertainty band, the naive weighted average (neglecting all correlations) $\bar{Y}_\text{NWA}$ is shown as a dashed gray line and the most deviating combination obtained by decorrelation $\bar{Y}'$ is shown as a dotted green line.}
    \label{fig:toy_model_data}
    \vspace{-0.4cm}
\end{figure*}
A `local' decorrelation (inspired by \cite{Leplumey:2025kvv}) can be defined by
\begin{equation}\label{eq:local_decorr}
    f^\text{L}_{ij}(\alpha^\text{L},\mathcal{E}) = \exp\Biggl[-\left(\frac{1-\alpha^\text{L}}{\alpha^\text{L}}\times\frac{x_i-x_j}{\mathcal{E}}\right)^2\Biggr],
\end{equation}
parameterized by $\alpha^\text{L}$. Here the $x_{i,j}$ are the abscissas of the input data and $\mathcal{E}$ is a (dimensionful) scaling which can be absorbed into $\alpha^\text{L}$. Values of $\mathcal{E}$ are chosen in this work to clearly display changes in the spectra and $\mathcal{M}$ across the full range $\alpha^\text{L}$. Other choices of $\mathcal{E}$ only change the scaling of $\alpha^\text{L}$ and do not change $\max\mathcal{M}$.

Applying the decorrelation models of Eq.~\eqref{eq:global_decorr} and Eq.~\eqref{eq:local_decorr} to the covariance matrices of the input data combined to produce the spectrum $O_i$ generates a modified spectrum, $\tilde O_i$, which serves as input to $\mathcal{M}$. Maximization of $\mathcal{M}$ then determines the deviating spectra, from which covariance matrices are evaluated corresponding to the new correlation-strength systematic uncertainty on $O_i$. These, in turn, propagate to uncertainties on derived quantities. A more detailed study could consider individual decorrelation of different systematic uncertainties (variation of $\alpha_{\text{sys,}m}$ parameters). However, at least in the global case, this is already effectively achieved. Assuming approximately constant relative contributions to the total systematic uncertainty, variation of each $\alpha_{\text{sys,}m}^\text{G}$ is covered by variation of $\alpha_{\text{sys}}^\text{G}$ in a limited range. Hence a more detailed consideration of this effect is not strictly necessary.

In the local decorrelation procedure, for $\alpha^\text{L}$ decreasing, correlations are reduced more between data points with a greater separation. Such a decorrelation is physically motivated by considering that correlations may only exist over a limited range, e.g.\ due to changing experimental conditions or analyses, and further that different systematics dominate in different regions of spectra. Additionally this can simulate effects such as a detector re-calibration between points, leading to different correlations in different regions of data. In the context of dispersive HVP, this would potentially apply to measurements by direct scan experiments, such as CMD-3~\cite{CMD-3:2023alj, CMD-3:2023rfe}. Such effects are in principle automatically accounted for if input covariance matrices from experimental analyses are sufficiently detailed.

\subsection{Toy model demonstration}

The use of the measure and decorrelation procedures to generate an additional systematic uncertainty is demonstrated using a simple toy model in which two datasets $d^{(1,2)}$ are input to a combination, shown in Fig.~\ref{fig:toy_model_data}. These consist of two data points each (indexed as $d_{1,2}$), with constant values $Y=0.9,\,1.1$ respectively. Both have a systematic (statistical) uncertainty of $10\%$ (1\%); they are taken to be totally statistically uncorrelated and are assumed uncorrelated with one another. Their default systematic correlation coefficients $\rho_{ij}^0$ are indicated on the right panel of Fig.~\ref{fig:toy_model_data}; their magnitudes are assumed to be imperfectly known. The datasets are to be combined into a single description, comprising two bins (clusters) at $X\sim0.875$ and $X\sim1.125$, in line with KNTW procedures described elsewhere~\cite{keshavarzi:2018mgv,keshavarzi:2019abf}.

\begin{figure}[!t]
    \centering
    \includegraphics[width=\linewidth]{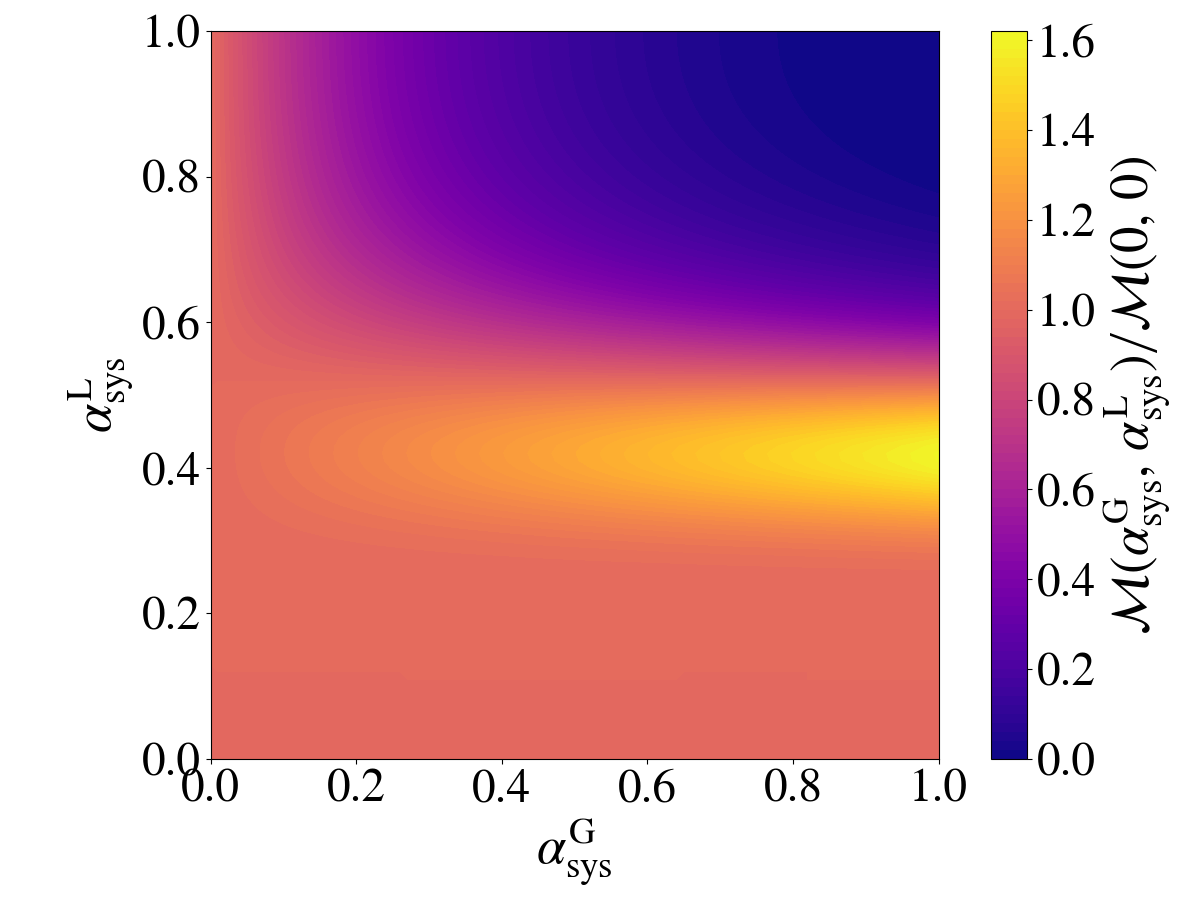}
    \caption{Plot of the measure $\mathcal{M}$ as defined in Eq.~\eqref{eq:measure} for the toy model under consideration. Global decorrelation parameterized by $\alpha_\text{sys}^\text{G}$ and local decorrelation parameterized by $\alpha_\text{sys}^\text{L}$ are performed in accordance with Eqs.~\eqref{eq:global_decorr} and~\eqref{eq:local_decorr}. The measure is plotted normalized to its value at $\alpha_\text{sys}^\text{G}=\alpha_\text{sys}^\text{L}=0$.}
    \label{fig:toy_model_variation}
    \vspace{-0.4cm}
\end{figure}

Using the KNTW combination procedure with full account of the given correlations, the identical result $\bar Y$ for both clusters is $\bar Y = 0.986\pm0.008_\text{stat}\pm0.070_\text{sys}$, which is $<1$ due to the $\rho^0_{ij}<1$ values in the $d^{(1)}$ block. To estimate the additional correlation strength uncertainty on this, the decorrelation procedures according to Eqs.~\eqref{eq:global_decorr} and~\eqref{eq:local_decorr} are applied.
Fig.~\ref{fig:toy_model_variation} shows the values of the measure $\mathcal{M}$ as $\alpha^\text{G}_\text{sys}$ and $\alpha^\text{L}_\text{sys}$ are varied simultaneously. The subscript `sys' here and in the remainder of the paper indicates that systematic correlations alone are varied. 

\begin{figure}[!t]
    \centering
    \includegraphics[width=\linewidth]{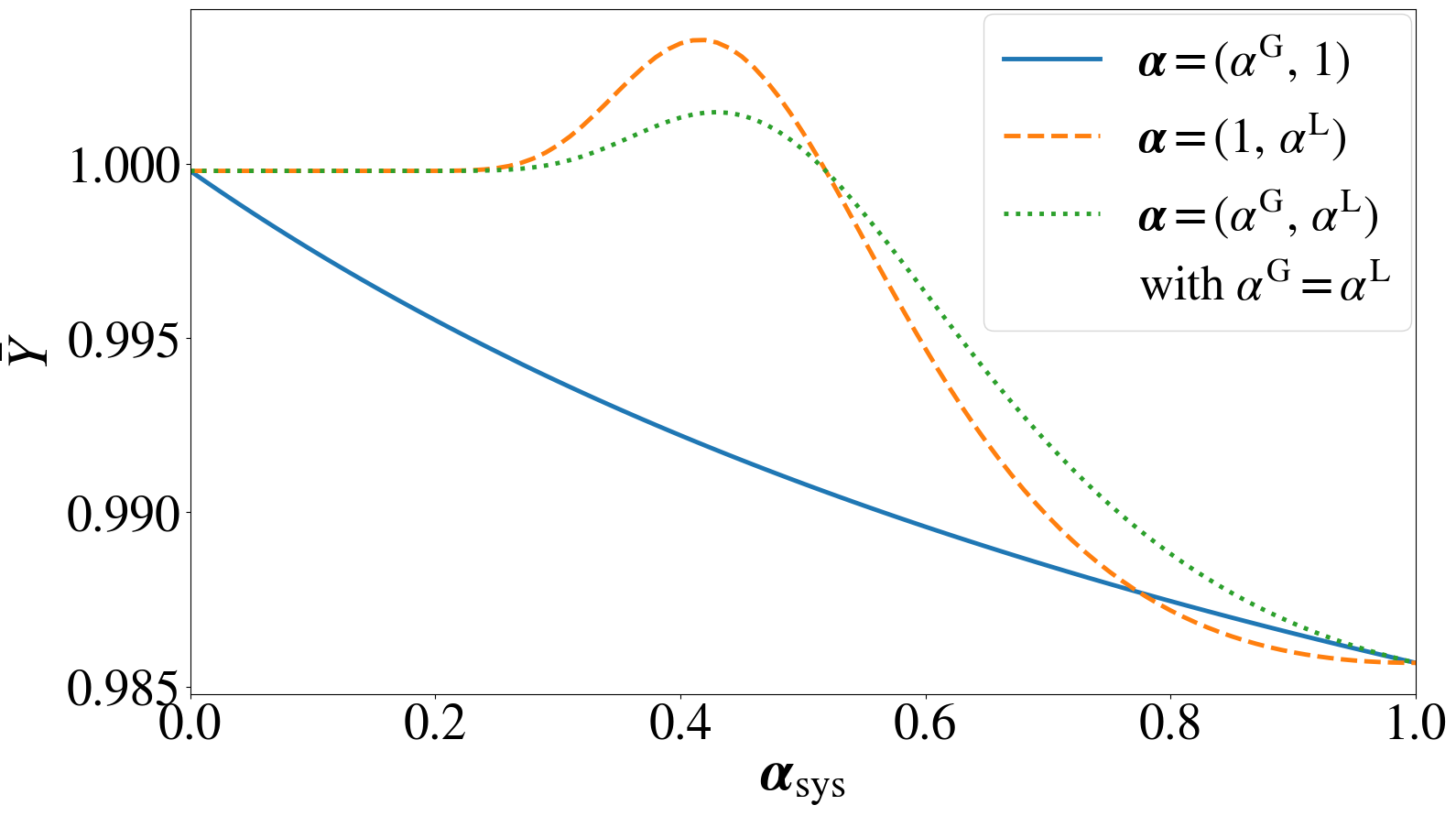}
    \caption{Change in the averaged cross section for particular choices of decorrelation from Eq.~\eqref{eq:global_decorr} and~\eqref{eq:local_decorr}. The scenarios are specified in the legend and correspond to horizontal, vertical and diagonal slices from $(1, 1)$ in Fig.~\ref{fig:toy_model_variation}.}
    \label{fig:toy_model_slices}
    \vspace{-0.4cm}
\end{figure}

As $\alpha^\text{G}_\text{sys}$ decreases, global decorrelation occurs and the off-diagonal entries are reduced by an equal factor $\alpha^\text{G}_\text{sys}$ such that the datasets are considered on increasingly even footing. This change is monotonic and ends with $\bar Y =1$ for each cluster; the expected naive average. This is consistent with the change in $\bar Y(\alpha^\text{G}_\text{sys})$ shown in Fig.~\ref{fig:toy_model_slices}.
When locally decorrelating, the correlation between $d^{(2)}_{1,2}$ is reduced more rapidly than that between $d^{(1)}_{1,2}$, so $d^{(2)}$ is favored in the fit for particular local decorrelations. This leads to $\bar Y>1$ and hence a greater maximum deviation of the spectrum (and $\max\mathcal{M}$) than can be achieved with global decorrelation alone. When mixing global and local decorrelations (e.g. the dotted green line), the change in the spectrum and measure is less pronounced as the discrepancy between the correlations of the datasets does not grow to the same extent.

Taking $\max\mathcal{M}$ at $\boldsymbol{\alpha}=(1.00,\,0.40)$, the additional uncertainty $d^\rho\bar Y = 0.018$ derived by our procedure aligns with the maximal change in $\bar Y$:
\begin{equation}\label{eq:toy_model_drho}
    \bar Y = 0.986\pm0.008_\text{stat}\pm0.070_\text{sys}\pm0.018_\rho \, ,
\end{equation}
where this $\rho$ uncertainty alone results in $\bar{Y}$ being consistent with 1. The new uncertainty is larger in this circumstance than the statistical uncertainties but almost negligible compared to the systematic uncertainties, marginally increasing the total uncertainty from $7.1\%$ to $7.4\%$. Had the input systematic uncertainties been smaller, this uncertainty would be more relevant. The additional uncertainty quantified in Eq.~\eqref{eq:toy_model_drho} is clearly dependent on the chosen default correlation assumption. For example, had one started at no correlation (i.e. $\bar Y=1$), the greatest discrepancy would be with our default and a smaller uncertainty would be quoted.

\section{APPLICATION TO THE KNT19 COMBINATION}\label{application}

The procedure introduced here to quantify the impact of differing correlation treatments in data combinations and to provide reliable estimates of the associated systematic uncertainties has been developed to be incorporated as a new component of the next full KNTW update. However, as the KNTW combination is currently subject to software blinding~\cite{keshavarzi:2024bli}, in the following section the method is applied to the KNT19 combination~\cite{keshavarzi:2019abf}, using the decorrelation models defined in Eq.~\eqref{eq:global_decorr} and Eq.~\eqref{eq:local_decorr} for the systematic correlations used in KNT19. This provides a detailed test case that both preserves the KNTW blinding and enables a direct exploration of the results that entered WP20~\cite{Aoyama:2020ynm}. 

Systematic uncertainties arising from the use of correlations are quantified in each channel by finding the largest deviations in $\mathcal{M}$ when applying global (Eq.~\eqref{eq:global_decorr}) and local (Eq.~\eqref{eq:local_decorr}) decorrelation procedures to systematic correlations. For local decorrelation, $\mathcal{E}=250$~MeV is chosen. From Eq.~\eqref{eq:Crho}, the procedure described above induces additional covariance matrices on the combined spectrum, which are subsequently propagated to integrated quantities that take the hadronic cross section as input. The observables considered here are those reported in the KNT19 update: the leading order HVP contributions to the anomalous magnetic moments of the electron ($a_e^\mathrm{HVP}$), muon ($a_\mu^\mathrm{HVP}$), and tau ($a_\tau^\mathrm{HVP}$), to the ground-state hyperfine splitting of muonium ($\Delta\nu_\mathrm{Mu}^\mathrm{HVP}$), and the five-flavor hadronic contribution to the running of the QED coupling evaluated at the $Z$ boson mass ($\Delta\alpha_\mathrm{had}^{(5)}(M_Z^2)$). 

The resulting correlation-strength systematic uncertainties on these quantities, $d^\rho$, are shown in Table~\ref{tab:errors_table}. While these additional uncertainties contribute at the level of the total uncertainty, they are not generally dominant. Their impact is most significant for $\Delta\alpha_\mathrm{had}^{(5)}(M_Z^2)$, whose integrand is more strongly weighted towards higher energies than the other observables, granting it a greater sensitivity to the higher-energy inclusive ($e^+e^- \rightarrow \text{all hadrons}$) channel above $\sim 2$ GeV. Importantly, this does not arise from an unusually large correlation-strength uncertainty in the inclusive channel itself, which is comparable to those of most other channels. Rather, it reflects the relatively small correlation-strength uncertainty in the important $\pi^+\pi^-$ channel, which dominates the lower-energy contributions to other observables and thereby suppresses their overall $d^\rho$ values. 
\begin{table*}
    \centering

        \begin{tabular}{cccccc}
        \hline\hline
        Channel            & $\ d^\rho a_e^\text{HVP}\times10^{14}\ $ & $\ d^\rho a_\mu^\text{HVP}\times10^{10}\ $ & $\ d^\rho a_\tau^\text{HVP}\times10^{8}\ $ & $\ d^\rho \Delta\alpha_\text{had}^{(5)}(M_Z^2)\times10^{4}\ $ & $\ d^\rho \Delta\nu_\text{Mu}^\text{HVP}\text{ (Hz)}\ $ \\\hline
        $\pi^+\pi^-$                   & 0.14 & 0.45 & 0.06 & 0.01 & 0.13\\
        $\pi^+\pi^-\pi^0$              & 0.15 & 0.58 & 0.28 & 0.07 & 0.20\\
        $\pi^+\pi^-2\pi^0$             & 0.07 & 0.30 & 0.21 & 0.07 & 0.11\\
        $2\pi^+2\pi^-$                 & 0.04 & 0.17 & 0.14 & 0.05 & 0.06\\
        $K^+K^-$                       & 0.16 & 0.61 & 0.29 & 0.06 & 0.21\\
        Inclusive $(\sqrt{s}\gtrsim2\text{ GeV})$                     
                                       & 0.08 & 0.34 & 0.50 & 0.81 & 0.15\\
        Total $d^\rho$ uncertainty              & 0.29 & 1.08 & 0.72 & 0.82 & 0.38\\ \noalign{\vskip 1.5mm}

        KNT19 uncertainty              & 0.66 & 2.42 & 1.39 & 1.12 & 0.82\\
        \noalign{\vskip 1.5mm}
        Updated uncertainty            & 0.72 & 2.65 & 1.57 & 1.39 & 0.90\\ 
        \% error contribution          & 16.2 & 16.6 & 21.1 & 34.9 & 17.7\\\hline\hline
    \end{tabular}
    \caption{Correlation strength uncertainties $d^\rho$ for the leading order HVP contributions to the lepton anomalous magnetic moments $a_{e,\mu,\tau}^\text{HVP}$, the running of the QED coupling at the $Z$ mass $\Delta\alpha_\text{had}^{(5)}(M_Z^2)$ and to the ground-state hyperfine splitting of muonium ($\Delta\nu_\mathrm{Mu}^\mathrm{HVP}$). The leading six channels are shown, as is the quadrature sum for all channels (the total $d^\rho$ uncertainty). The third-to-last row shows the total uncertainty for each observable determined in KNT19~\cite{keshavarzi:2019abf}. The second-to-last row gives the updated KNT19 uncertainty as the quadrature sum of the previous KNT19 uncertainty and new correlation-strength uncertainties. The final row shows the percentage contribution of this new uncertainty to the updated error budget.}
    \label{tab:errors_table}
    \vspace{-0.4cm}
\end{table*}

Although the full parameter space of global and local decorrelations can be scanned simultaneously, comparatively studying these decorrelation scenarios separately allows a better understanding of the impact of these procedures on the combined spectra. In all but one (subleading) channel, the maximum deviation of $\alpha^\text{G}_\text{sys}$ occurs at full decorrelation.\footnote{\label{ft}The exception, $K^+K^-\pi^+\pi^-$, has $\max\mathcal{M}$ at $\alpha^\text{G}_\text{sys}=0.02$, with a $\sim0.1\%$ excess. Since this has no impact on the quoted uncertainty for any integrated observable, this edge case is neglected and for all practical purposes the most extreme deviation can be considered to occur at $\alpha^\text{G}_\text{sys}=0$.} For most channels -- including $\pi^+\pi^-$ -- $\mathcal{M}(\alpha)$ monotonically increases with decreasing $\alpha_\text{sys}^\text{G}$. For local decorrelation, nine channels in the KNT19 combination have a maximum deviation at $\alpha_\text{sys}^\text{L}\neq0$. For these, $\max\mathcal{M}$ compared to $\mathcal{M}(0)$ is negligible or inconsequential. This is because $\max\mathcal{M}$ either results from regions where $\sigma\sim0$ or the channels' overall contributions are small. For example, in the $\pi^+\pi^-\pi^0$ channel, $\mathcal{M}$ is maximized at $\alpha_\text{sys}^\text{L}=0.07$ due to a region directly above the $\phi$ resonance where the cross section is suppressed. In all nine cases, the impact on the resultant uncertainty compared to $\alpha_\text{sys}=0$ is irrelevant at the quoted precision. Equivalent results for $\max\mathcal{M}$ are found when scanning the full parameter space of $\alpha_\text{sys}^\text{G}$ and $\alpha_\text{sys}^\text{L}$.
\begin{figure}[!t]
    \centering
    \includegraphics[width=\linewidth]{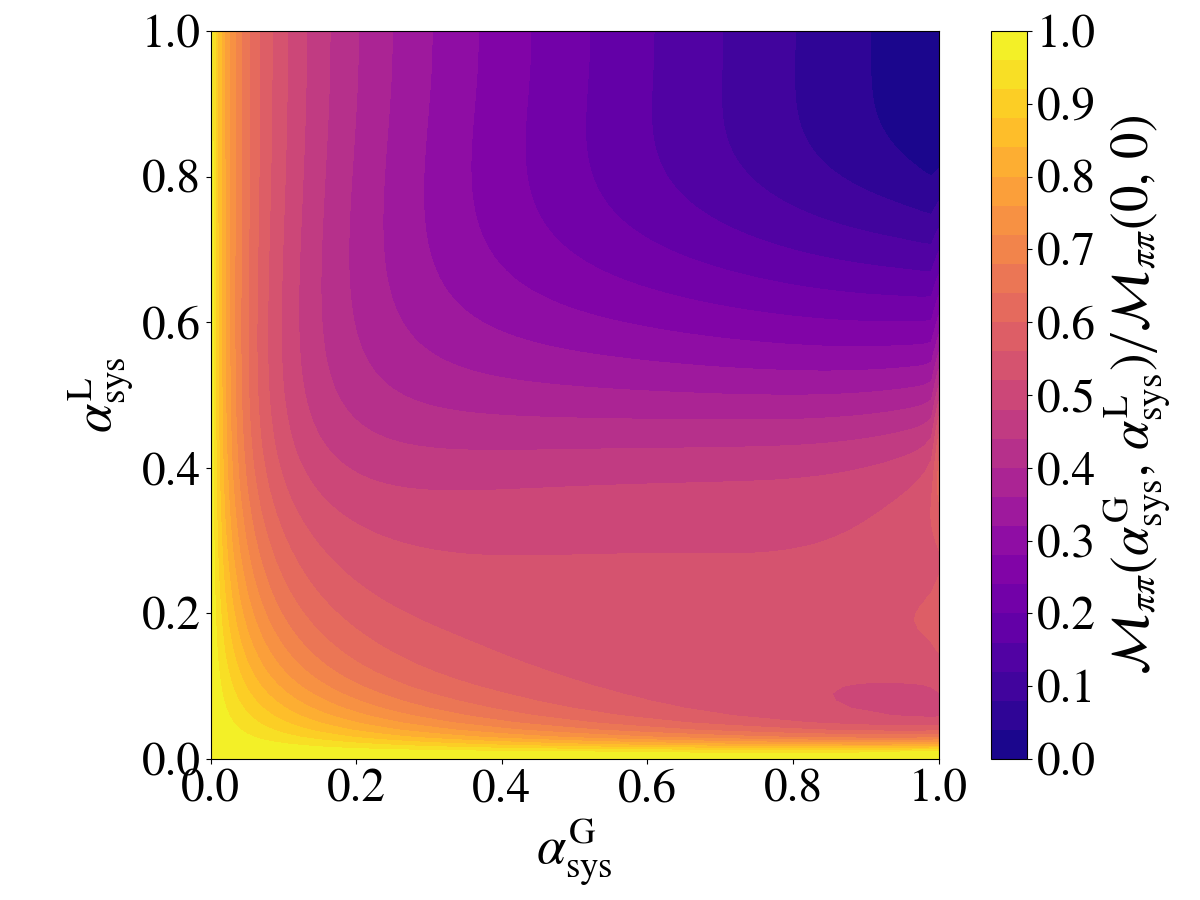}
    \caption{As Fig.~\ref{fig:toy_model_variation}, but for the $\pi^+\pi^-$ channel.}
    \label{fig:measure_sys_2D}
    \vspace{-0.4cm}
\end{figure}
\begin{figure}[!t]
    \centering
    \includegraphics[width=\linewidth]{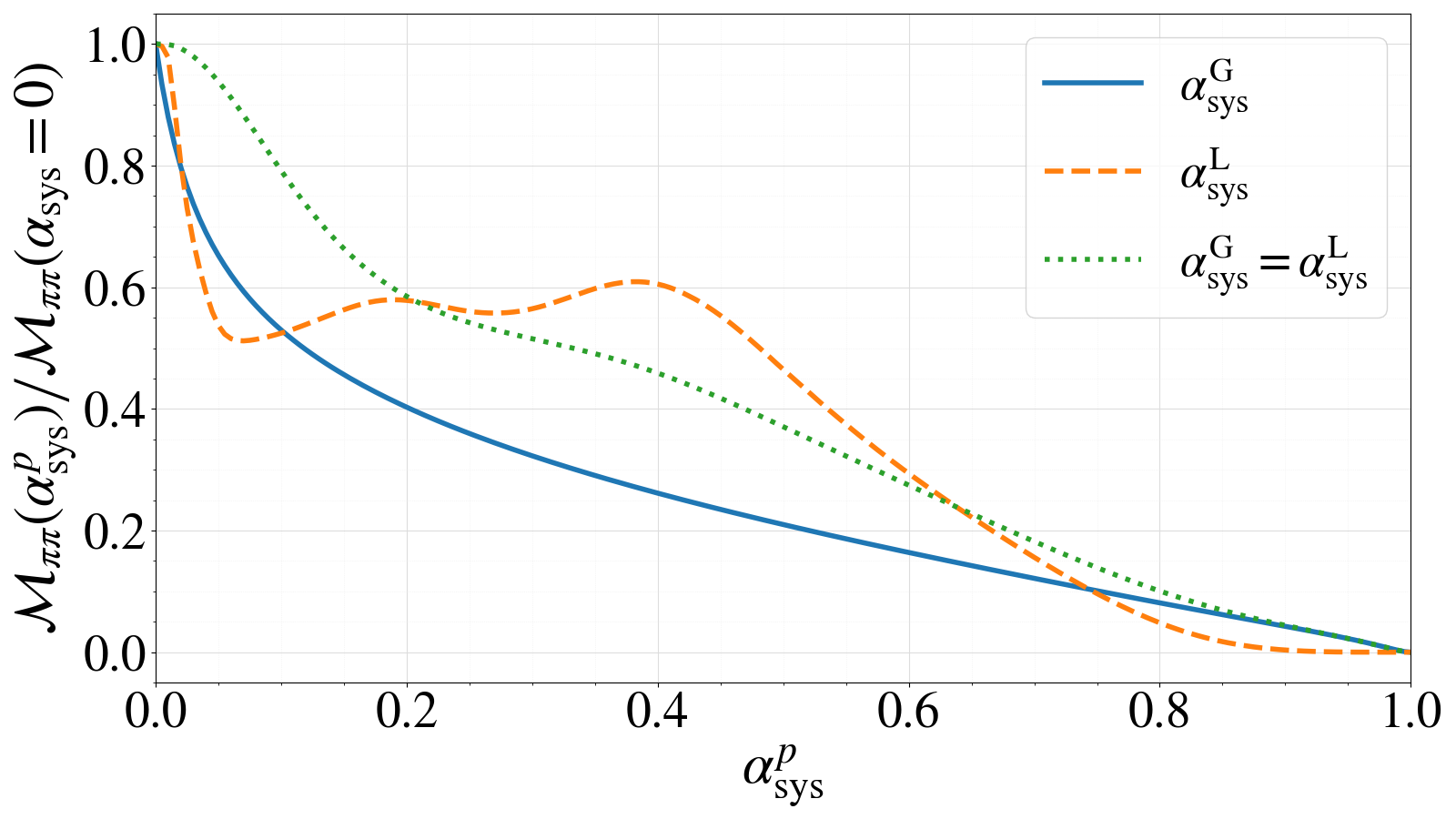}
    \caption{Values of the measure $\mathcal{M}$ defined in Eq.~\eqref{eq:measure} for variation of $\alpha_\text{sys}^\text{G, L}$ (as defined in Eqs.~\eqref{eq:global_decorr} and~\eqref{eq:local_decorr}) in the KNT19 $\pi^+\pi^-$ data combination. Curves are normalized to the common $\mathcal{M}_{\pi\pi}(\alpha_\text{sys}^\text{G}=0)=\mathcal{M}_{\pi\pi}(\alpha_\text{sys}^\text{L}=0)$.}
    \label{fig:measure_sys}
    \vspace{-0.4cm}
\end{figure}

\subsection{The $\pi^+\pi^-$ channel}

Fig.~\ref{fig:measure_sys_2D} and Fig.~\ref{fig:measure_sys} show the values of the measure $\mathcal{M}$ as $\alpha^\text{G}_\text{sys}$ and $\alpha^\text{L}_\text{sys}$ are varied for the $\pi^+\pi^-$ channel in KNT19. The result is the same for both simultaneous and separate parameter scans with $\max\mathcal{M}$ consistently found at $\alpha_\text{sys}=0$. The resulting channel correlation strength uncertainty is $d^\rho a_\mu^{\pi^+\pi^-}=0.45\times10^{-10}$. 
\begin{figure}[!t]
    \centering
    \includegraphics[width=\linewidth]{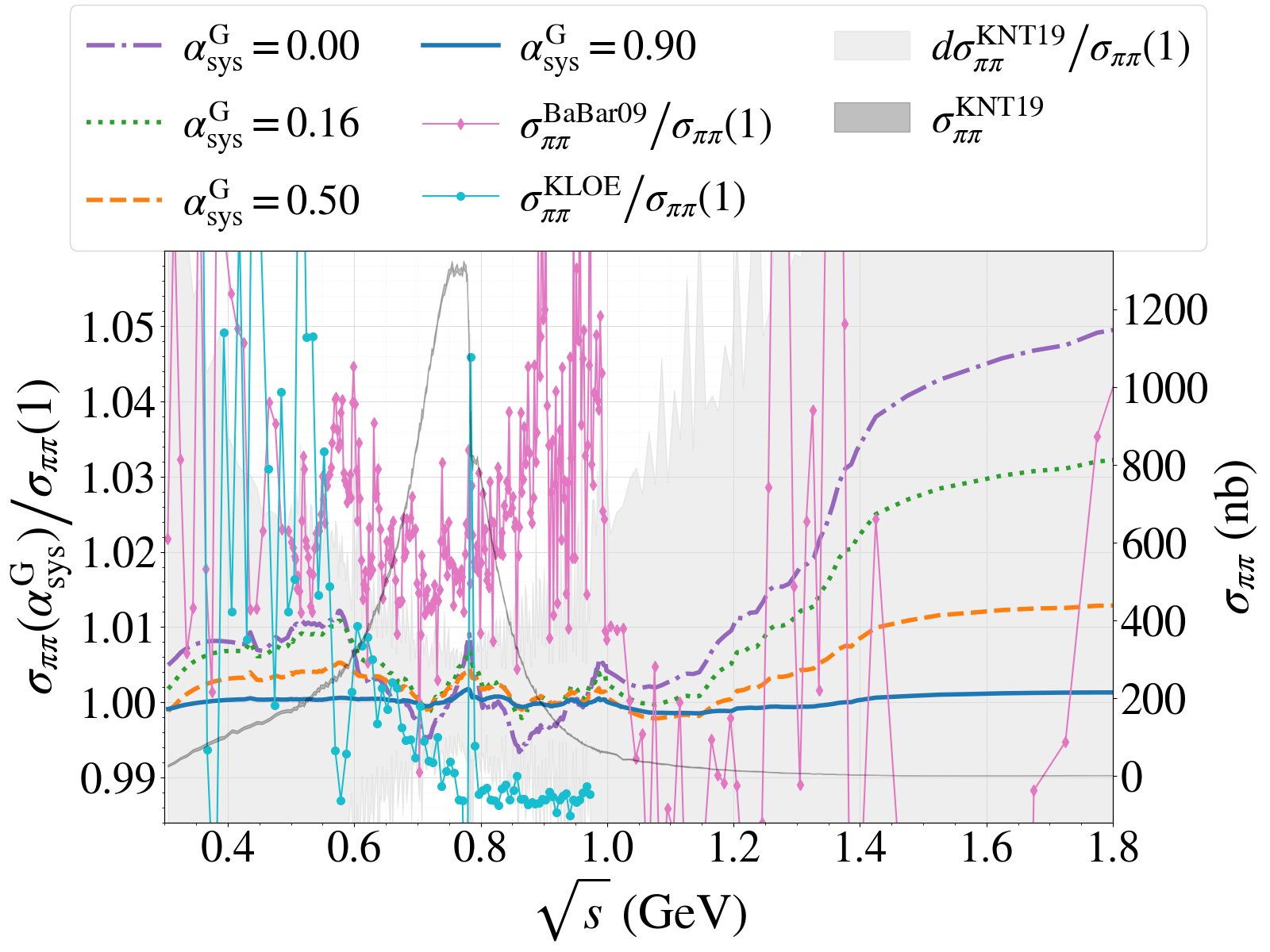}
    \caption{Specific global decorrelations of the KNT19 $\pi^+\pi^-$ spectrum $\sigma_{\pi\pi}^\text{KNT19}$ compared to the KNT19 uncertainty $d\sigma_{\pi\pi}^\text{KNT19}$~\cite{keshavarzi:2019abf}. These, and spectra of $\sigma_{\pi\pi}$ from various measurements, are shown normalized to $\sigma_{\pi\pi}^\text{KNT19}=\sigma_{\pi\pi}(1)$ (left axis). For readability, statistical fluctuations of the combinations (but not the data) are smoothed using a Savitsky-Golay~\cite{doi:10.1021/ac60214a047} filter. The BaBar~\cite{BaBar:2009wpw, BaBar:2012bdw} and KLOE~\cite{KLOE:2008fmq, KLOE:2010qei, KLOE:2012anl, KLOE-2:2017fda} data are shown as pink and cyan lines with diamond and circle markers respectively. The purple dot-dashed, green dotted, orange dashed and solid blue lines correspond to values of $\alpha_\text{sys}^\text{G}$ as in Eq.~\eqref{eq:global_decorr}. The normalized KNT19 $\sigma_{\pi\pi}$ errors are overlaid as a gray band. The KNT19 two pion spectrum is shown as a black band (right axis).}
    \label{fig:global_decorr_spectra}
\end{figure}

Fig.~\ref{fig:global_decorr_spectra} shows spectra for specific global decorrelations of the KNT19 $\pi^+\pi^-$ spectrum $\sigma_{\pi\pi}^\text{KNT19}$, compared to the KNT19 uncertainty $d\sigma_{\pi\pi}^\text{KNT19}$. For large $\alpha^\text{G}_\text{sys}$, deviations from the default $\sigma_{\pi\pi}^\text{KNT19}$ remain small. Despite growing with decreasing $\alpha^\text{G}_\text{sys}$, they do not become inconsistent with $d\sigma_{\pi\pi}^\text{KNT19}$ at any energy. Corroborating Figs.~\ref{fig:measure_sys_2D} and~\ref{fig:measure_sys}, Fig.~\ref{fig:global_decorr_spectra} also demonstrates that the measure is maximized at $\alpha^\text{G}_\text{sys}=0$, i.e.\ full decorrelation.

\begin{figure}[!t]
    \centering
    \includegraphics[width=\linewidth]{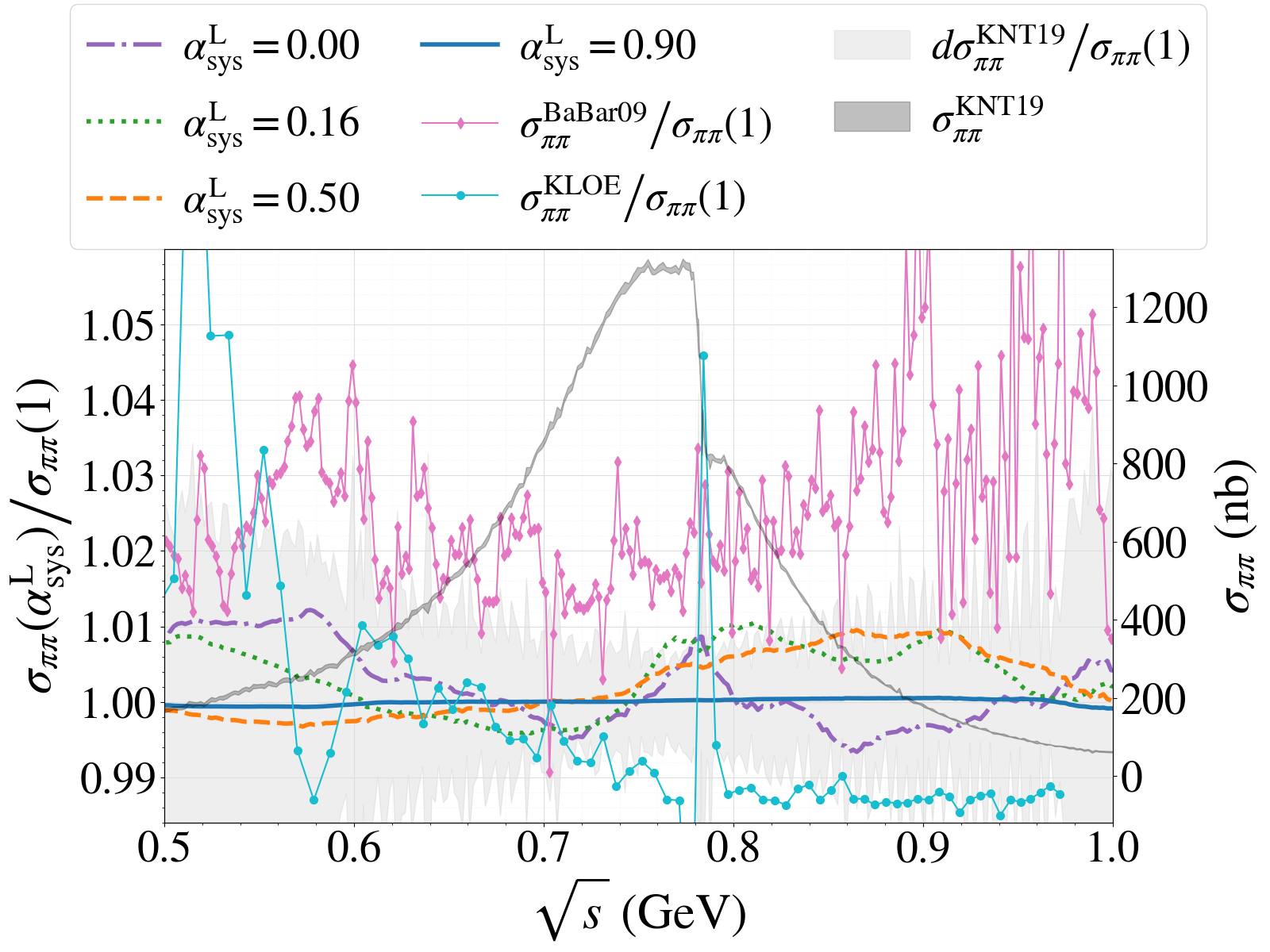}
    \caption{Specific local decorrelations of the KNT19 $\pi^+\pi^-$ spectrum. As Fig.~\ref{fig:global_decorr_spectra}, except $\alpha_\text{sys}^\text{L}$ (Eq.~\eqref{eq:local_decorr}) is varied rather than $\alpha_\text{sys}^\text{G}$ and data are shown in the relevant restricted range.}
    \label{fig:local_decorr_spectra}
    \vspace{-0.4cm}
\end{figure}

For local decorrelation -- shown in Fig.~\ref{fig:local_decorr_spectra} -- although the most extreme overall deviation in $\sigma_{\pi\pi}$ again corresponds to full decorrelation, this is less obvious than in Fig.~\ref{fig:global_decorr_spectra}, as the local case shows a more complicated situation than the global. The local decorrelations in Fig.~\ref{fig:local_decorr_spectra} are shown in a restricted range around the $\rho$-resonance as significant differences from Fig.~\ref{fig:global_decorr_spectra} are only observed there. Unlike for global decorrelation, where spectra with larger $\alpha^\text{G}_\text{sys}$ resemble the spectrum with $\alpha_\text{sys}^\text{G}=0$, significantly different lineshapes are observed for $\alpha_\text{sys}^\text{L}\neq0,\,1$. This is expected from the two local maxima and minima in $\mathcal{M}$($\alpha^\text{L}_\text{sys}$) in Fig.~\ref{fig:measure_sys} and could result from a non-trivial structure of the channel's covariance matrix when isolating increasingly narrow regions. 


\vspace{0.2cm}
\subsubsection{Discussion of $d^\rho a_\mu^{\pi^+\pi^-}$}
\vspace{0.1cm}

\begin{figure}
    \centering
    \includegraphics[width=0.49\textwidth]{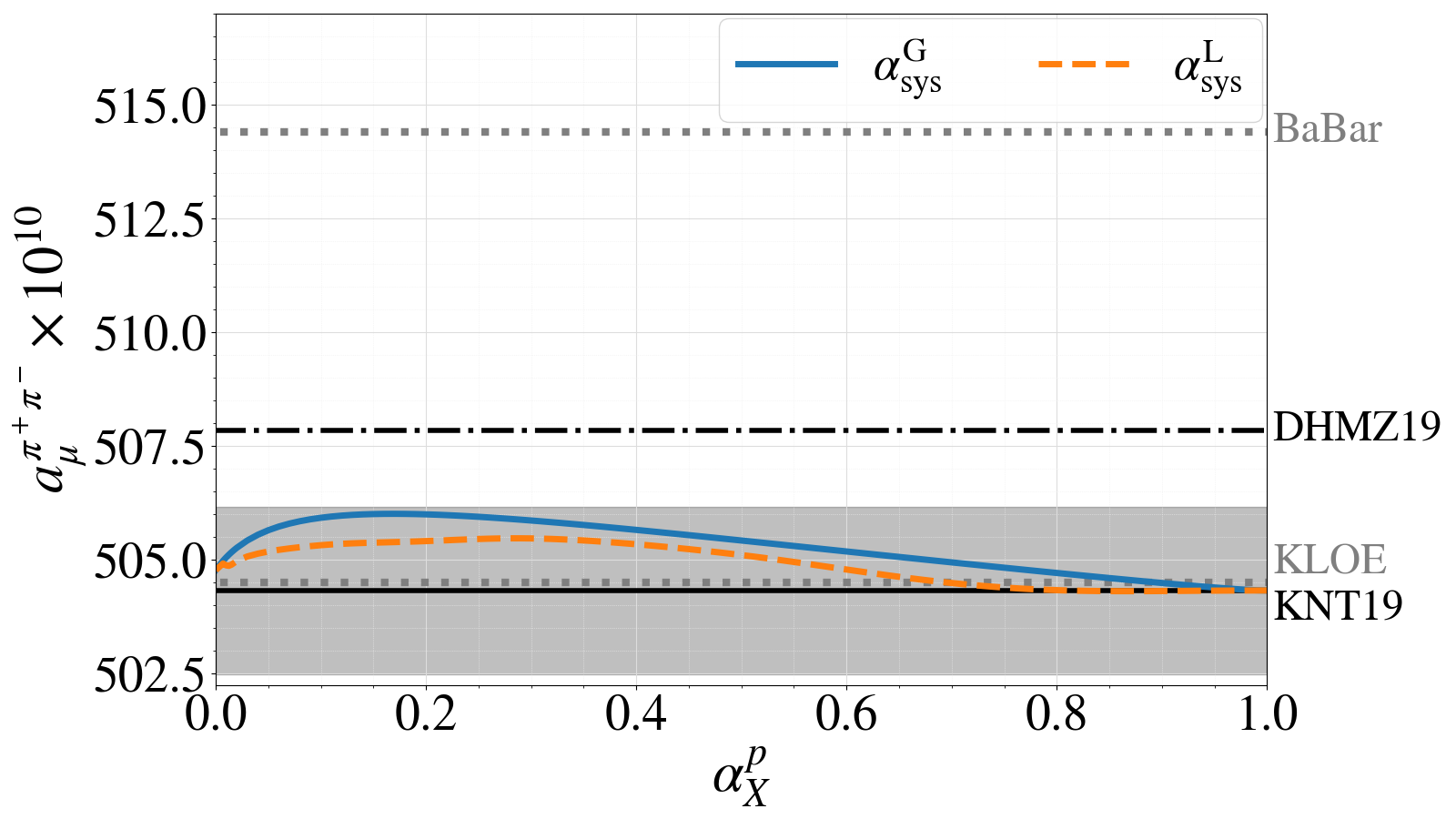}
    \caption{Contribution of the KNT19 $\pi^+\pi^-$ channel to the muon anomaly from scans over $\alpha$ parameters given in the legend and detailed in the text. The KNT19 value of $a_\mu^{\pi^+\pi^-}$, and shaded error band, are from~\cite{keshavarzi:2019abf}, and the comparison DHMZ19 value is from~\cite{davier:2019can}, which uses the same data. The BaBar~\cite{BaBar:2009wpw, BaBar:2012bdw} and KLOE~\cite{KLOE:2008fmq, KLOE:2010qei, KLOE:2012anl, KLOE-2:2017fda} predictions are shown as dotted lines.}
    \label{fig:amu_impacts}
    \vspace{-0.4cm}
\end{figure}

Although all uncertainties on derived quantities have been determined at the level of the underlying spectra, a discussion at the level of the integrated observable $a_\mu^{\pi^+\pi^-}$ is warranted. Fig.~\ref{fig:amu_impacts} shows the change in $a_\mu^{\pi^+\pi^-}$ for separate global and local decorrelations of $\sigma_{\pi\pi}^\text{KNT19}$. This shows that, in contrast to $\max\mathcal{M}$ occurring at full decorrelation, the maximum deviation of $a_\mu^{\pi^+\pi^-}$ occurs at $\alpha_\text{sys}^\text{G}=0.16$.\footnote{This still lies within the original uncertainty band of the KNT19 result (shaded gray in Fig.~\ref{fig:amu_impacts}).} If we were to adopt this as an estimate of the correlation strength uncertainty, $d^\rho a_\mu^{\pi^+\pi^-}$ would increase to $1.68\times10^{-10}$. This can be understood from Fig.~\ref{fig:global_decorr_spectra}. When $\alpha^\text{G}_\text{sys}$ is decreased to 0.16, $\sigma_{\pi\pi}$ predominantly increases or remains constant. At $\alpha^\text{G}_\text{sys}=0.16$, an on-average three-per-mille increase of the spectrum occurs, causing a similar increase in $a_\mu^{\pi^+\pi^-}$. When $\alpha^\text{G}_\text{sys}$ decreases further, local decreases in $\sigma_{\pi\pi}$ -- at $\sim0.7$ GeV and $\sim0.9$ GeV -- become sizable and increase $\mathcal{M}$ (and the point-to-point uncertainty). However, as the former energy is the peak of the $\rho$-resonance, the resulting cancellations in the $a_\mu^{\pi^+\pi^-}$ integral are significant. Consequently, using $\alpha_\text{sys}^\text{G}=0.16$ to estimate the $\pi^+\pi^-$ systematic uncertainty would correspond to reduced and potentially underestimated uncertainties on the spectrum. It would also lead to an overall inconsistent treatment of correlation strength uncertainties across different channels, where in all other relevant channels $\max\mathcal{M}$ gives maximized deviations in derived quantities. 

\vspace{0.2cm}
\subsubsection{Comparison with DHMZ19}\label{sec:amu}
\vspace{0.1cm}

At the time of KNT19 (and, therefore, WP20), the most precise and thus dominant datasets in the $\pi^+\pi^-$ channel were BaBar~\cite{BaBar:2009wpw, BaBar:2012bdw} and KLOE~\cite{KLOE:2008fmq, KLOE:2010qei, KLOE:2012anl, KLOE-2:2017fda}. These datasets exhibit a tension where the BaBar/KLOE $\sigma_{\pi\pi}$ data were (before the CMD-3 measurements) generally the highest/lowest lying data in the central $\rho$ resonance region, with BaBar exceeding KLOE by $\sim3\sigma$. The BaBar and KLOE $\sigma_{\pi\pi}$ spectra are shown normalized to $\sigma_{\pi\pi}^\text{KNT19}$ in Figs.~\ref{fig:spectra_comparison},~\ref{fig:global_decorr_spectra} and~\ref{fig:local_decorr_spectra}.

Fig.~\ref{fig:amu_impacts} highlights how the KNT19 combination favors a more KLOE-like result, whereas the DHMZ19 result exhibits a relatively stronger influence from BaBar. This difference has historically been understood to predominantly arise from the different treatment of correlations in the respective fits. As described in Sec.~\ref{sec:corr_comb}, the KNTW approach maximally utilizes all available correlation information, resulting in a $\pi^+\pi^-$ data combination that is more strongly constrained by correlations than in DHMZ. It was therefore expected that the decorrelation procedure implemented here would reproduce, to some extent, WP20's dominant “BaBar–KLOE” systematic uncertainty of $2.8\times10^{-10}$~\cite{Aoyama:2020ynm}. Although this uncertainty was derived at the level of the integrated observable rather than from differences between the combined $\pi^+\pi^-$ spectra, it was introduced to account for the difference between BaBar and KLOE, and the difference $a_\mu^{\pi^+\pi^-}[\text{DHMZ19]} - a_\mu^{\pi^+\pi^-}[\text{KNT19]} = 3.6\times10^{-10}$.

However, $d^\rho a_\mu^{\pi^+\pi^-}=0.45\times10^{-10}$ is approximately six times smaller than WP20's “BaBar–KLOE” uncertainty. In fact, it is eight times smaller than the DHMZ19/KNT19 difference. Even the observable-based estimate of $1.68\times10^{-10}$ remains approximately $40\%-50\%$ smaller. This clearly demonstrates that variation of the systematic correlations alone does not reproduce the “BaBar–KLOE” systematic uncertainty. An explanation for this discrepancy is provided in Appendix~\ref{Appendix}.

\vspace{0.2cm}
\subsection{The $K^+K^-$ channel}\label{sec:KK}
\vspace{0.1cm}
A large correlation strength systematic uncertainty is obtained in the $K^+K^-$ channel. Studying local decorrelation reveals that this is dominated by effects in the $\phi$ resonance region. This channel also exhibits the largest $\max\mathcal{M}$, which is understood to drive the sizable uncertainties ($2$–$2.8$ systematic standard deviations) in all integrated observables shown in Table~\ref{tab:errors_table}. These effects are driven by the tension between the dominant BaBar~\cite{BaBar:2013kkx} and CMD-3~\cite{CMD3:2018kkx} measurements in the $K^+K^-$ channel. In the $\phi$ resonance region, correlations in the BaBar data are relatively weak compared to the near fully correlated CMD-3 data, leading to BaBar being favored in the KNTW fit. When these correlations are neglected, a result closer to the simple weighted average of the two is obtained. Notably, the differences between these $K^+K^-$ datasets are comparable to those observed in the $\pi^+\pi^-$ channel. A similar effect may therefore be expected there when the CMD-3 $\pi^+\pi^-$ data are included in the next full KNTW update.

\section{CONCLUSIONS}\label{sec:conclusions}

In this work, we have developed a general and systematic framework to quantify uncertainties arising from assumptions about correlations in data combinations. While the treatment of correlated uncertainties is a longstanding and essential component of precision measurements, imperfect knowledge of correlations -- specifically those associated with systematic uncertainties -- introduces an additional layer of uncertainty that is typically not evaluated in a systematic manner. This issue is especially relevant in dispersive determinations of the hadronic vacuum polarization contributions to observables such as the muon anomalous magnetic moment, where the combination of highly correlated $e^+e^- \rightarrow \mathrm{hadrons}$ data plays a central role.

The procedure introduced here provides a well-defined means to explore variations in the assumed systematic correlation structure through physically motivated decorrelation scenarios, and to identify the maximal deviation of the resulting combined spectrum using a generalized measure, $\mathcal{M}$. In doing so, it enables the construction of correlation-strength systematic uncertainties and associated covariance matrices directly at the level of the combined spectrum, rather than inferring such effects from differences in derived quantities. This represents a significant conceptual and practical improvement over previous approaches, such as those adopted in the 2020 White Paper from the Muon $g$$-$$2$ Theory Initiative~\cite{Aoyama:2020ynm}, where additional systematic uncertainties were introduced in an ad hoc manner based on differences between integrated results. By construction, the present framework retains sensitivity to local spectral variations and provides a consistent and reproducible method to propagate these effects to derived observables.

Application of this method to the KNT19 data combination~\cite{keshavarzi:2019abf} demonstrates several important features. The resulting correlation-strength uncertainties are generally subdominant but non-negligible contributions to the total uncertainty budget of integrated observables such as $a_\mu^\mathrm{HVP}$ and $\Delta\alpha_\mathrm{had}^{(5)}(M_Z^2)$. More broadly, the framework provides a quantitative means to disentangle the effects of correlation assumptions from intrinsic discrepancies between datasets. In particular, it clarifies the role of correlation treatments in shaping differences between existing data combinations, such as those of KNTW~\cite{keshavarzi:2024bli,keshavarzi:2018mgv,keshavarzi:2019abf} and DHMZ~\cite{Davier:2010nc, Davier:2010rnx, davier:2017zfy, davier:2019can, Davier:2023fpl}. Importantly, the method shows that variations in systematic correlations alone do not reproduce previously assigned uncertainties in key channels such as $\pi^+\pi^-$, including the dominant “BaBar–KLOE” uncertainty in WP20. This demonstrates that such uncertainties cannot be attributed solely to imperfectly known systematic correlations and their use in data fits, but instead arise from a more complex interplay between dataset tensions and methodological choices (see Appendix~\ref{Appendix}).

The definition of the uncertainty through the maximal deviation of the spectrum, obtained via $\mathcal{M}$, is conservative with respect to the resulting cross section uncertainties. However, this approach may be overly conservative should experiments provide quantitative constraints on the magnitude of systematic uncertainties and their correlations. In such cases, the range of variation of the decorrelation parameters may be restricted accordingly, yielding a more precise estimate of the associated uncertainty. More generally, the availability of detailed information on the correlations of individual systematic uncertainty sources from experimental collaborations would enable the construction of more accurate covariance matrices at the outset, and reduce the reliance on conservative assumptions in data combination procedures.

The procedure developed here is general and applicable beyond the specific case of $e^+e^- \rightarrow \mathrm{hadrons}$ data combinations. It will form an integral part of the next-generation KNTW analysis framework, where it will be applied to updated combinations, including forthcoming high-precision measurements. In particular, its application will be essential in assessing the impact of new data in tension-dominated channels such as $\pi^+\pi^-$. More broadly, progress in dispersive HVP studies will require both improved experimental measurements, with tensions quantified, and refined analysis methodologies, including the consistent treatment of correlated uncertainties. These developments are a necessary step to reconcile and potentially combine dispersive and lattice evaluations of $a_\mu^\mathrm{HVP}$, determine a single best estimate of $a_\mu^\mathrm{SM}$, and fully realize the potential of $a_\mu^\mathrm{exp}$.

\section*{ACKNOWLEDGMENTS}
The authors thank the Muon $g$$-$$2$ Theory Initiative -- in particular Fedor Ignatov, Peter Stoffer, Hartmut Wittig and the DHMZ group (Michel Davier, Andreas Hoecker, Bogdan Malaescu, Zhiqing Zhang) for useful discussions. They are also grateful to MITP for hosting the recent conference on the fine structure constant.
AK is supported by The Royal Society (URF$\backslash$R1$\backslash$231503). 
TT is supported by the STFC Consolidated Grant ST/X000699/1. AW is supported by a PGR studentship jointly funded by STFC and the Leverhulme Trust under LIP-2021-01. 

\appendix
\section{STATISTICAL CORRELATIONS IN $\pi^+\pi^-$} \label{Appendix}
\addcontentsline{toc}{section}{APPENDIX: $a_\mu^\text{HVP}$ \& IMPACTS OF STATISTICAL CORRELATIONS}

\begin{figure}
    \centering
    \includegraphics[width=\linewidth]{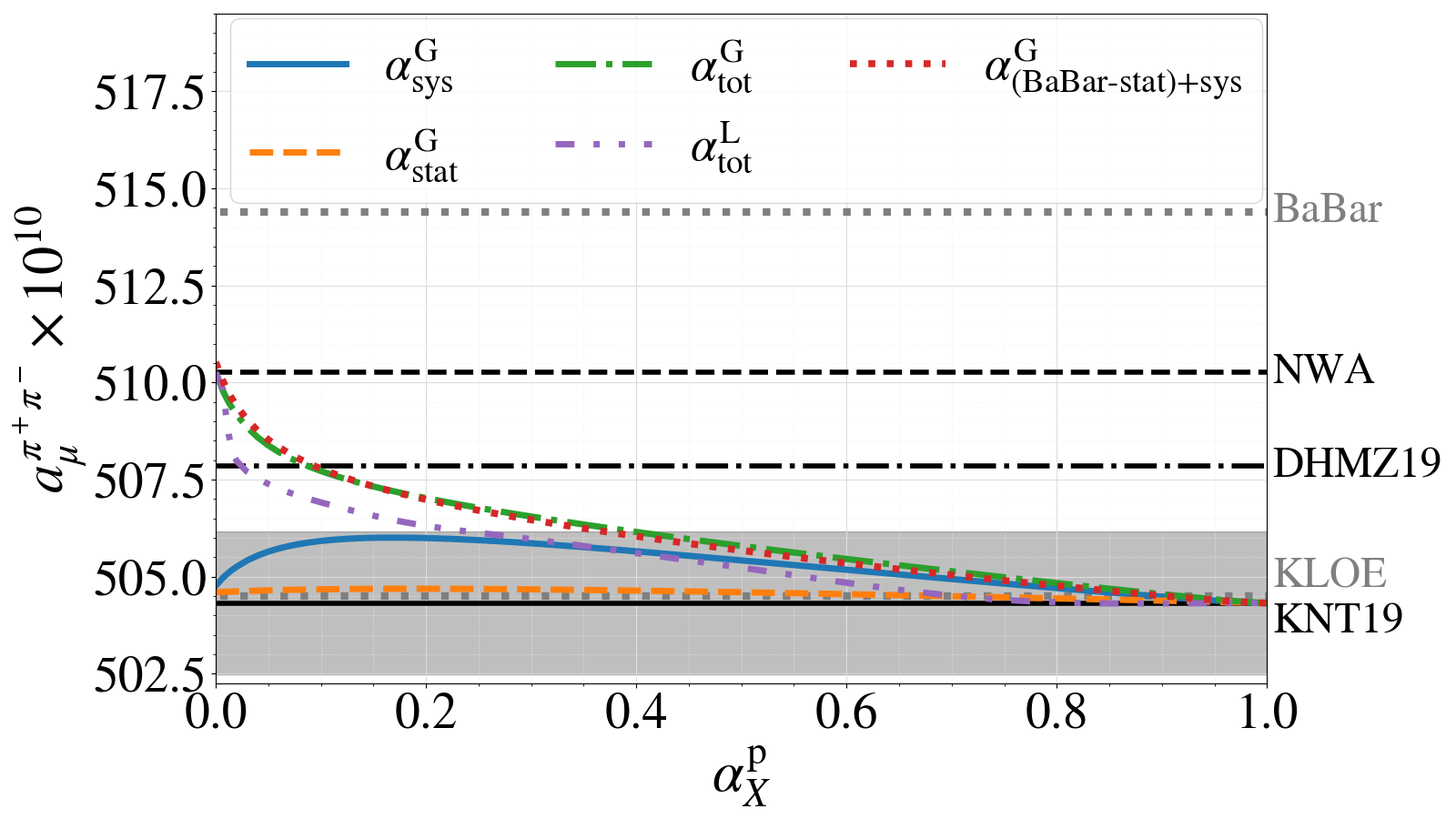}
    \caption{As Fig.~\ref{fig:amu_impacts} but including additional decorrelation scans. Here, `tot' denotes total, i.e. varying all (statistical and systematic correlations) simultaneously. The naive weighted average (NWA) value is produced neglecting all correlations.}
    \label{fig:amu_all}
\end{figure}

Whilst it is understood that differing treatments of correlated uncertainties are a principal source of the differences between the DHMZ and KNTW results, it is clear from Sec.~\ref{sec:amu} that varying the systematic correlations -- either by maximizing $\mathcal{M}$ or by maximizing the difference in the leading order HVP contribution $a_\mu^{\pi^+\pi^-}$ -- is insufficient to fully account for the observed difference of $a_\mu^{\pi^+\pi^-}[\text{DHMZ19]} - a_\mu^{\pi^+\pi^-}[\text{KNT19}] = 3.6\times10^{-10}$. A naive weighted average (NWA) of the KNT19 $\sigma_{\pi\pi}$ datasets below $1.8~\mathrm{GeV}$ -- where each bin is weighted solely by its local uncertainties -- yields $a_\mu^{\pi^+\pi^-}[\text{KNT19,\,NWA}] = 510.3(2.9)\times10^{-10}$, as shown in Fig.~\ref{fig:amu_all}. By construction, the NWA neglects all correlations (both statistical and systematic) -- denoted here by `tot' -- and corresponds to the limit $\alpha^\text{G,\,L}_\text{tot}=0$, as illustrated in Fig.~\ref{fig:amu_all}. Notably, $a_\mu^{\pi^+\pi^-}[\text{DHMZ19}]$ lies closer to this value than to $a_\mu^{\pi^+\pi^-}[\text{KNT19}]$. 

Although, as discussed in Sec.~\ref{sec:corr_comb}, statistical correlations are not subject to assumptions and associated uncertainties in the same way as systematic correlations, the behavior observed in the NWA suggests that they may nevertheless play a role in explaining the differences between the DHMZ and KNTW results. We therefore first consider the case in which only the $\sigma_{\pi\pi}$ statistical correlations are varied, while the systematic uncertainties are kept fixed. This is illustrated by the dashed orange line in Fig.~\ref{fig:amu_all}, which shows the effect on $a_\mu^{\pi^+\pi^-}[\text{KNT19}]$ when varying $C_\text{stat}$ via Eq.~\eqref{eq:global_decorr} for all $\pi^+\pi^-$ datasets in the KNT19 combination. Consistent results are obtained when applying local decorrelation (see Eq.~\eqref{eq:local_decorr}). However, since the $\sigma_{\pi\pi}$ systematic uncertainties are dominant, and only the BaBar, BESIII~\cite{BESIII:2015equ}, and KLOE datasets exhibit non-negligible statistical correlations, the impact of varying statistical correlations alone on both $a_\mu^{\pi^+\pi^-}$ and $\mathcal{M}$ is small. Consequently, in the following we restrict attention to variations of the total (statistical and systematic) correlations.

\begin{figure}
    \centering
    \includegraphics[width=\linewidth]{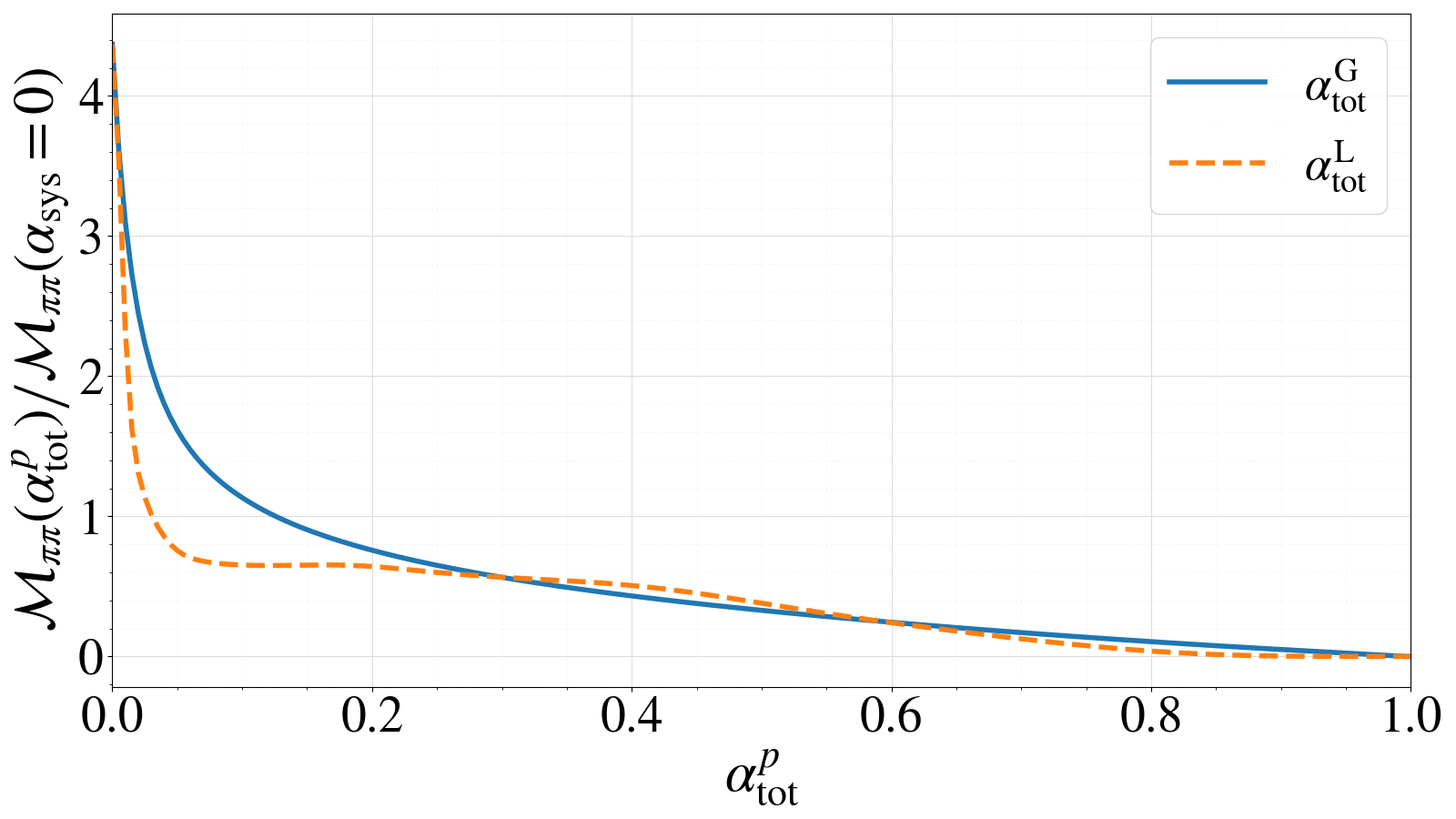}
    \caption{As Fig.~\ref{fig:measure_sys}, but with $\alpha^\text{G,\,L}_\text{tot}$ varied rather than $\alpha^\text{G,\,L}_\text{sys}$. The normalization is $\mathcal{M}_{\pi\pi}(\alpha_\text{sys}=0)$, as is used in Fig.~\ref{fig:measure_sys}, and not $\mathcal{M}_{\pi\pi}(\alpha_\text{tot}=0)$.}
    \label{fig:measure_tot}
    \vspace{-0.4cm}
\end{figure}\
\begin{figure}[!t]
    \centering
    \includegraphics[width=\linewidth]{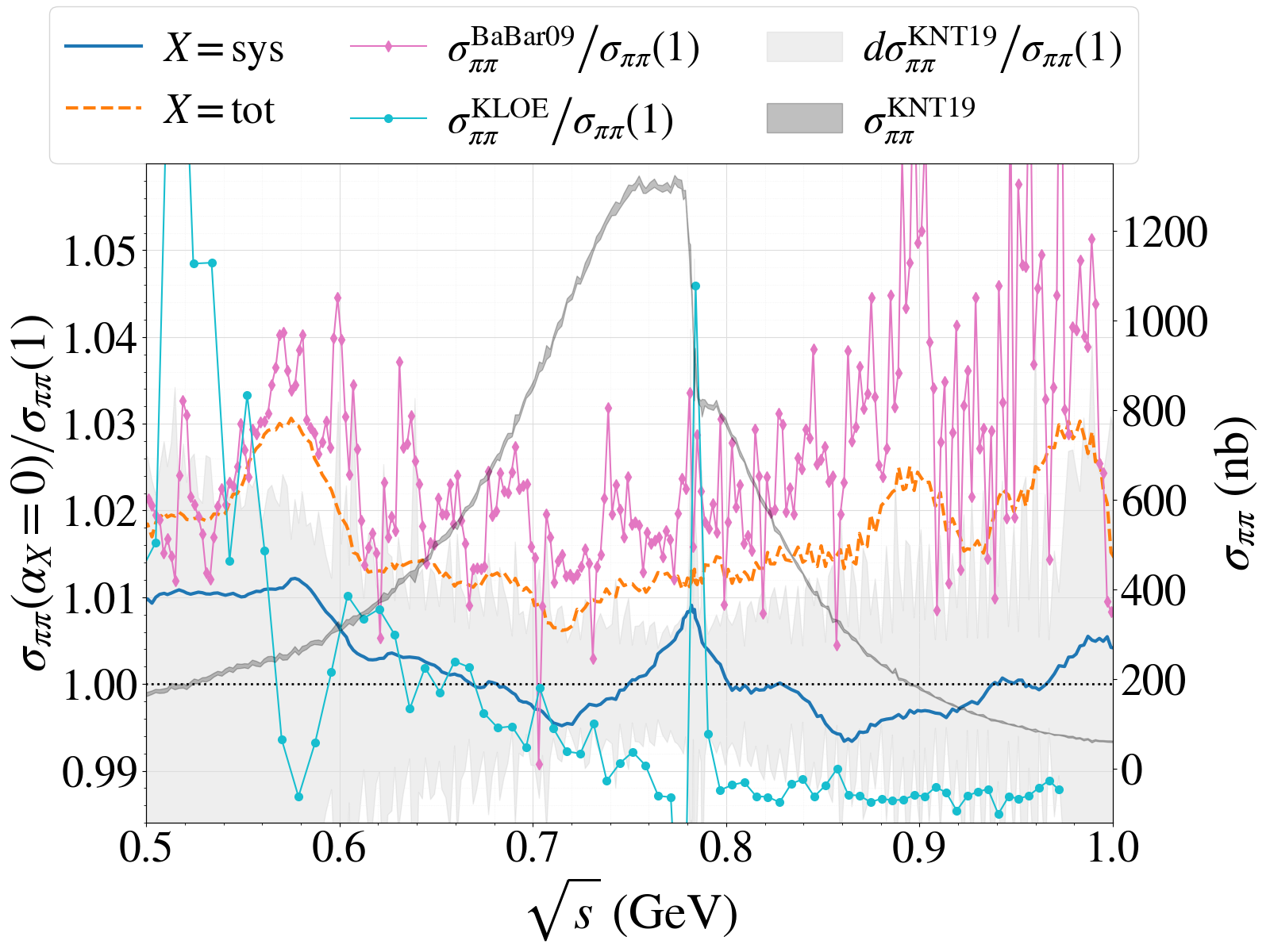}
    \caption{Plotted data are as in Fig.~\ref{fig:global_decorr_spectra}, except the orange dashed and solid blue lines correspond instead to values of $\alpha_\text{sys,tot}=0$, respectively.}
    \label{fig:both_decorr_spectra}
\end{figure}
In Fig.~\ref{fig:measure_tot}, the values of $\mathcal{M}$ for variations of $\alpha_\text{tot}^\text{G,L}$ are shown. To facilitate a direct comparison with Fig.~\ref{fig:measure_sys}, the same normalization factor is used for the ordinates. Significantly larger deviations in the spectra are observed when statistical correlations are decorrelated in addition to the systematic ones. This can also be seen in Fig.~\ref{fig:both_decorr_spectra}, where the fully decorrelated spectrum more closely resembles the higher-lying BaBar spectrum~\cite{BaBar:2009wpw, BaBar:2012bdw}. As a result, total decorrelation leads to values of $a_\mu^{\pi^+\pi^-}$ that are inconsistent with KNT19 for $\alpha_\text{tot}^{\text{G}}\lesssim0.40$ ($\alpha_\text{tot}^{\text{L}}\lesssim0.25$). Furthermore, $a_\mu^{\pi^+\pi^-}[\text{DHMZ19}]$ is reproduced for $\alpha_\text{tot}^{\text{G}}\sim0.08$ ($\alpha_\text{tot}^\text{L}\sim0.02$).

\begin{figure}[!t]
    \centering
    \includegraphics[width=0.95\linewidth]{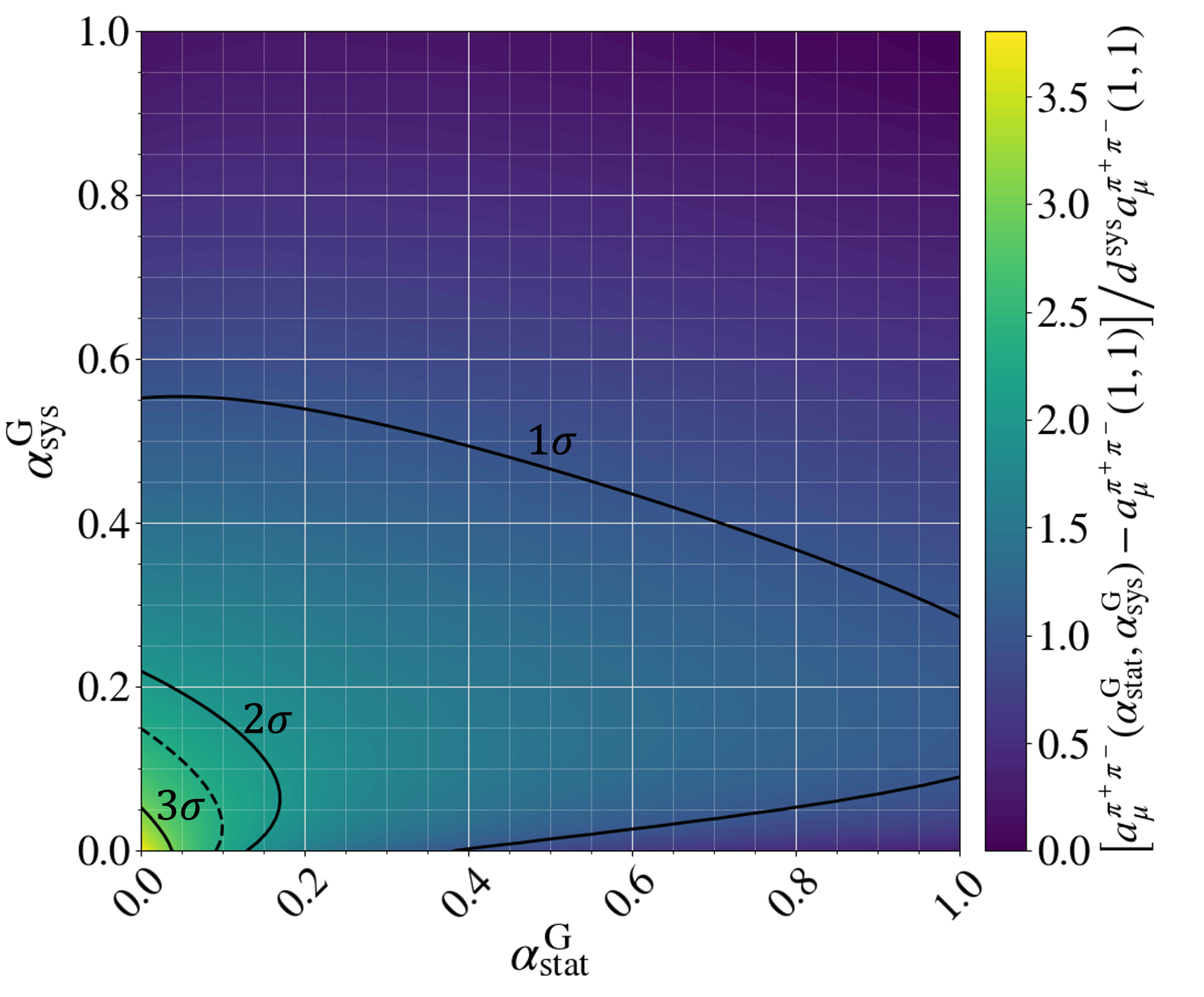}
    \caption{Differences between (KNTW framework) $a_\mu^{\pi^+\pi^-}$ values found by independently varying the statistical and systematic correlations (given by the scaling factors $\alpha^\text{G}_\text{stat}$ and $\alpha^\text{G}_\text{sys}$ respectively), normalized to the systematic uncertainty of KNT19. Labeled integer values of this ratio ($1\sigma$, $2\sigma$ and $3\sigma$) are indicated with solid black lines, and the DHMZ19 result~\cite{davier:2019can} is indicated by a dashed black line.}
    \label{fig:pipi_grid}
    \vspace{-0.4cm}
\end{figure}
The dependence of $a_\mu^{\pi^+\pi^-}$ on simultaneous global variations of statistical and systematic correlations, implemented through the decorrelation procedure
\begin{equation}
F_{ij}^{\pi\pi}(\alpha_\text{stat}^\text{G},\alpha_\text{sys}^\text{G}) = f_{ij}^\text{G}(\alpha_\text{stat}^\text{G}) \times f_{ij}^\text{G}(\alpha_\text{sys}^\text{G}) \, ,
\end{equation}
 is shown in Fig.~\ref{fig:pipi_grid} (see also Eqs.~\eqref{eq:decorr_factorised} and~\eqref{eq:global_decorr}). Significant deviations from $a_\mu^{\pi^+\pi^-}[\text{KNT19}]$ induced by simultaneous variation of statistical and systematic correlations are found only within a narrow region of parameter space centered around $(\alpha_\text{stat}^\text{G},\alpha_\text{sys}^\text{G}) = (0,0)$. The DHMZ result is also recovered in the vicinity of this region. This suggests that the DHMZ result is driven higher as their combination method may eliminate or reduce statistical as well as systematic correlations, despite statistical uncertainties and their correlations being well-defined features of data analyses.

\begin{figure*}[!t]
\centering
\text{BaBar Correlation Matrices}\\[0.0cm]
\vspace{-0.3cm}
\subfloat[Statistical\label{fig:BaBar_stat}]{
    \includegraphics[width=0.45\textwidth]{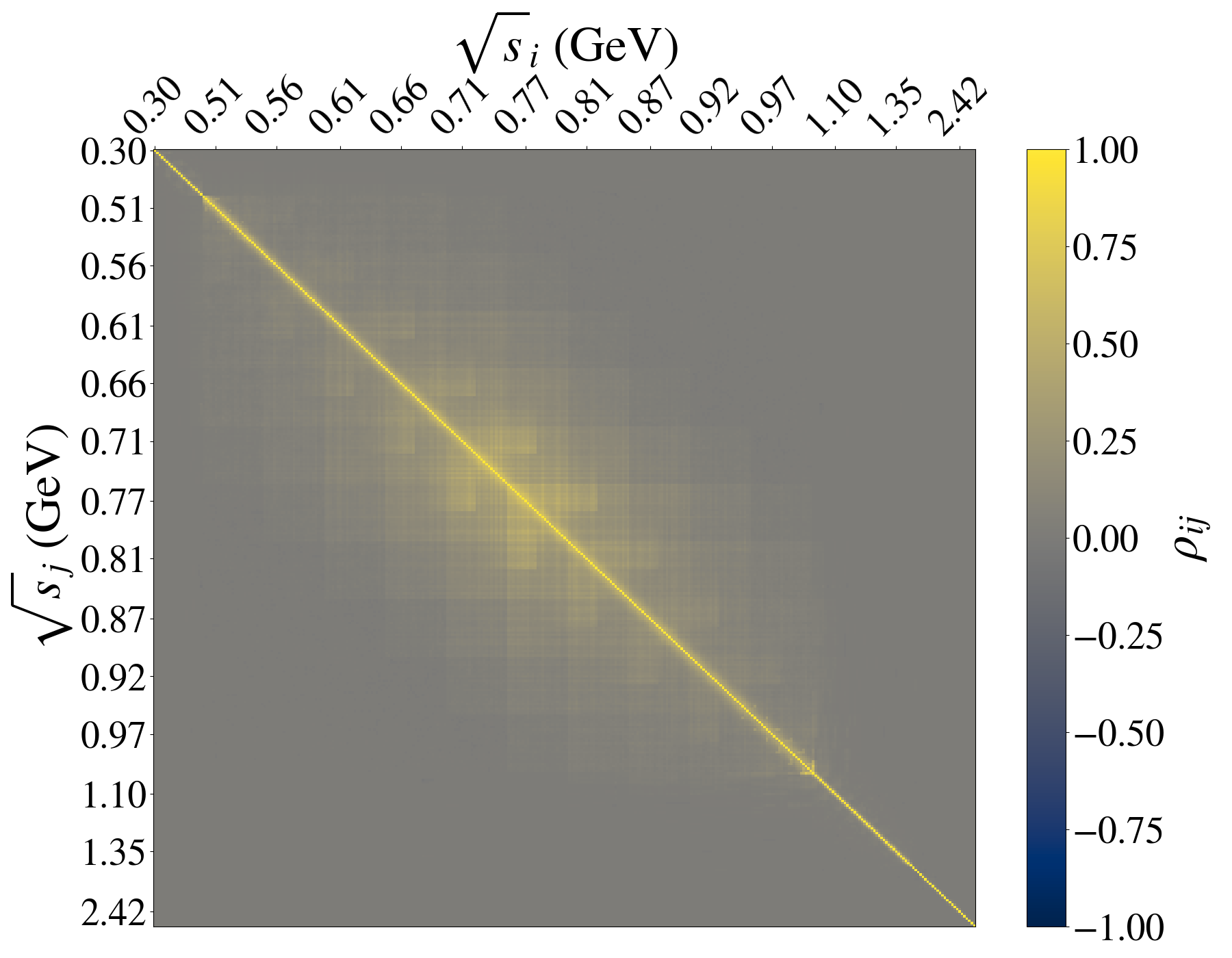}
}
\hspace{0.4cm}
\subfloat[Systematic\label{fig:BaBar_syst}]{
    \includegraphics[width=0.45\textwidth]{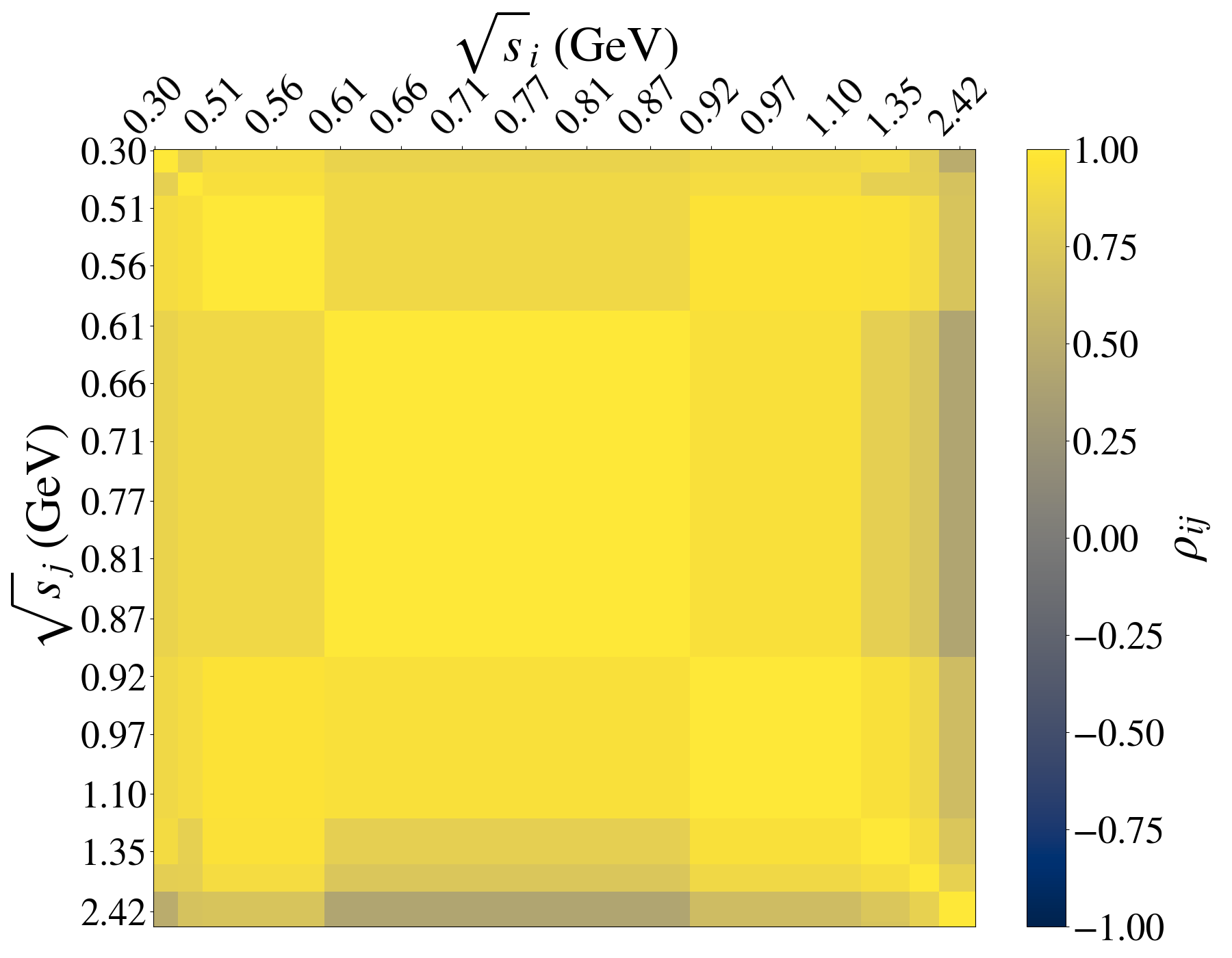}
}
\vspace{0.6cm}
\text{KLOE Correlation Matrices}\\[0.0cm]
\vspace{-0.3cm}
\subfloat[Statistical\label{fig:KLOE_stat}]{
    \includegraphics[width=0.45\textwidth]{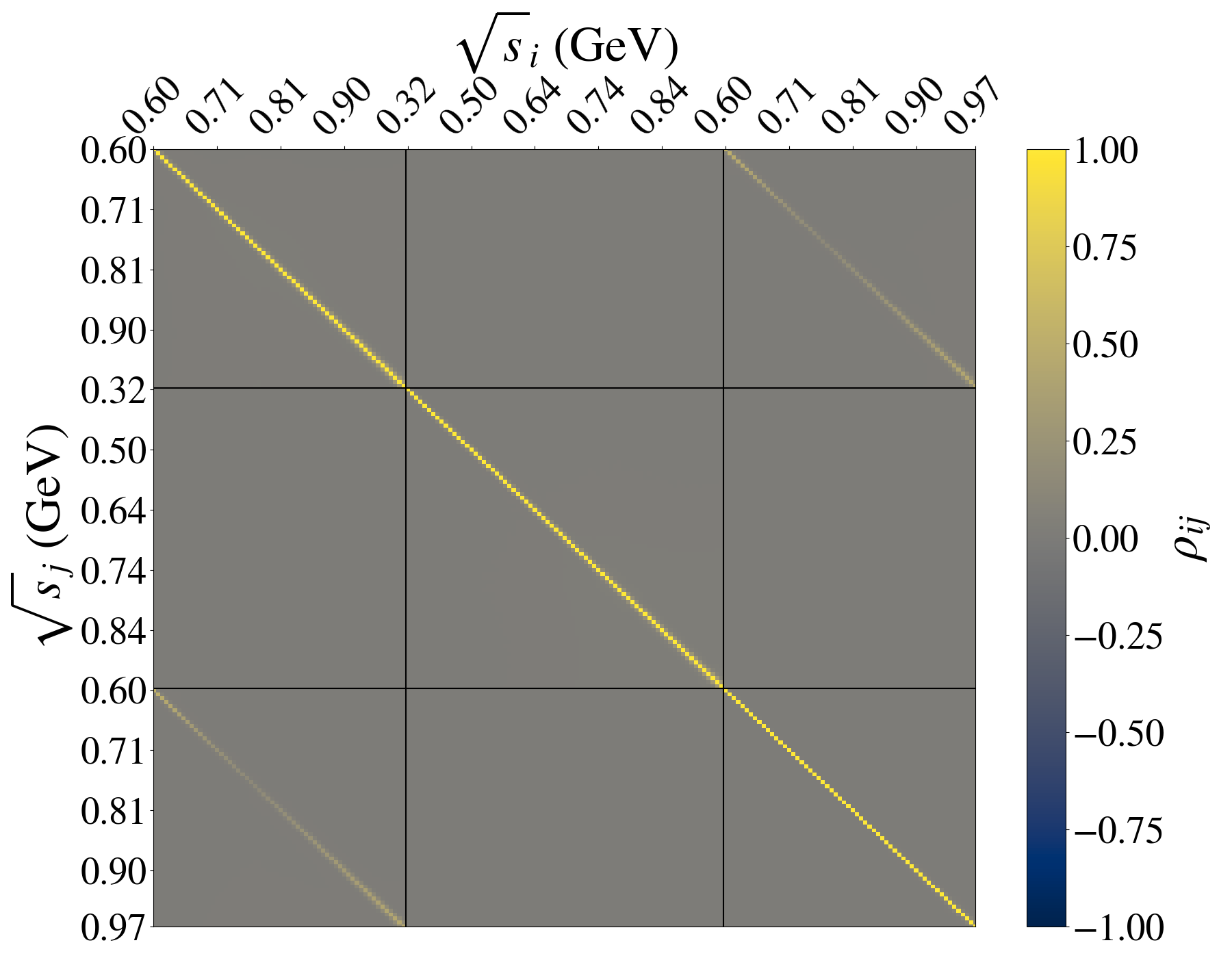}
}
\hspace{0.4cm}
\subfloat[Systematic\label{fig:KLOE_syst}]{
    \includegraphics[width=0.45\textwidth]{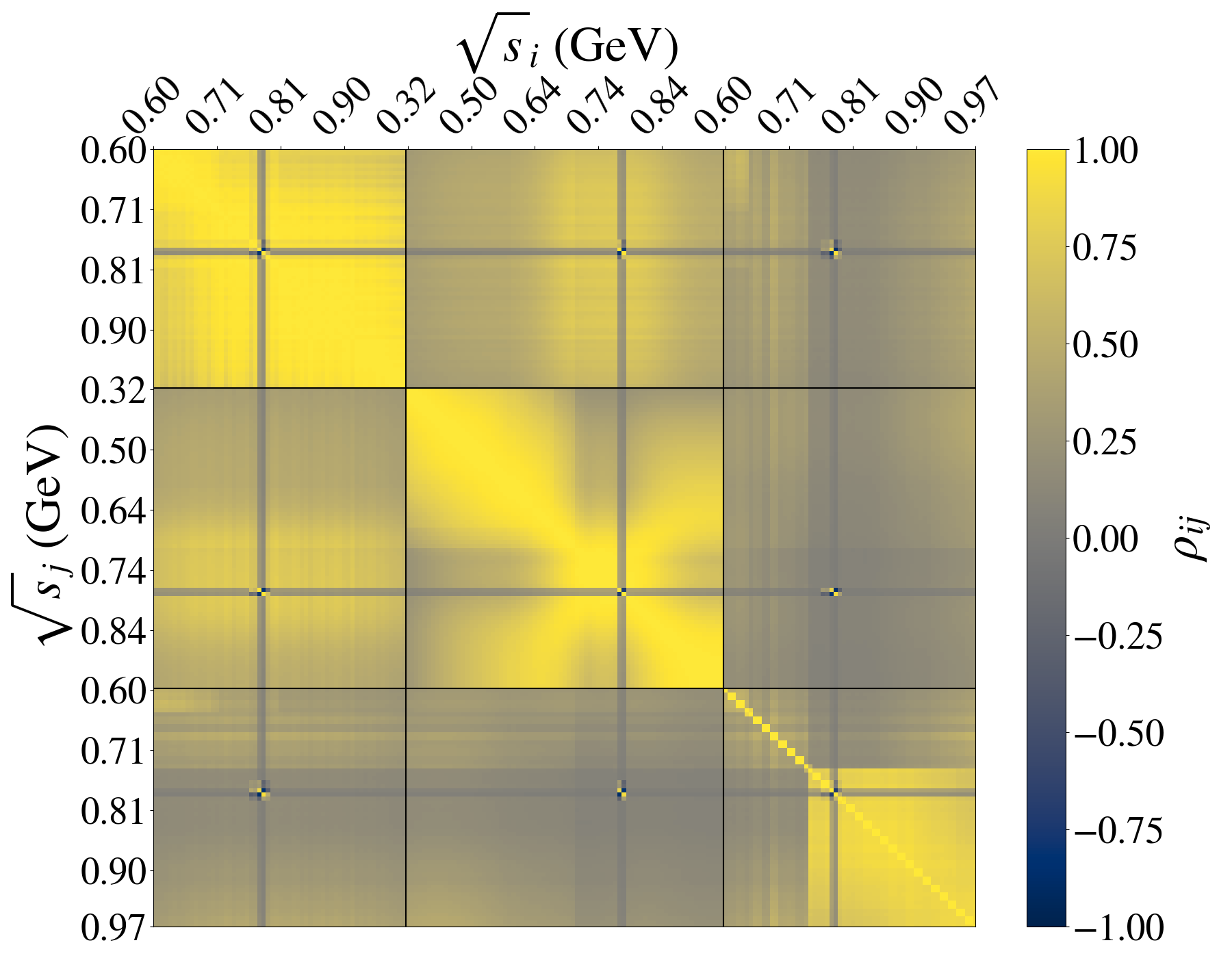}
}
\caption{
Correlation matrices for the $\pi^+\pi^-$ cross section measurements ($\sigma_{\pi\pi}$) from BaBar~\cite{BaBar:2009wpw, BaBar:2012bdw} and KLOE~\cite{KLOE:2008fmq, KLOE:2012anl, KLOE-2:2017fda, KLOE:2010qei}.
Vertical and horizontal black lines divide the KLOE matrices into blocks corresponding to the 2008~\cite{KLOE:2008fmq}, 2010~\cite{KLOE:2010qei}, and 2012~\cite{KLOE:2012anl} datasets.
}
\label{fig:cmats}

\vspace{-0.4cm}
\end{figure*}

The question that then arises is how modifying the influence of statistical correlations in $\sigma_{\pi\pi}$ leads the data combination to favor BaBar or KLOE, respectively. A particularly striking feature of Fig.~\ref{fig:amu_all} is the red dotted line, which illustrates the specific effect of decorrelating only the BaBar statistical correlations, together with all systematic correlations, in the KNT19 $\pi^+\pi^-$ dataset (with all other statistical correlations left unchanged). As can be seen, a behavior remarkably similar to that obtained when varying the full $C_\text{tot}$ is observed. This demonstrates that the difference between the DHMZ and KNT19 results arises predominantly from the reduction of BaBar statistical correlations in conjunction with the systematic correlations.\footnote{This is further supported by considering $\mathcal{M}(\alpha^\text{L}_\text{tot})$ versus $\mathcal{M}(\alpha^\text{L}_\text{sys})$, with reference to Fig.~\ref{fig:BaBar_stat} and Eq.~\eqref{eq:local_decorr}.}

The statistical and systematic correlation matrices derived from the covariance matrices provided by the BaBar~\cite{BaBar:2009wpw, BaBar:2012bdw} and KLOE~\cite{KLOE:2008fmq, KLOE:2012anl, KLOE-2:2017fda, KLOE:2010qei} experiments are illustrated in Fig.~\ref{fig:cmats}. Comparing Fig.~\ref{fig:BaBar_syst} and Fig.~\ref{fig:KLOE_syst}, BaBar is closer to full systematic correlation than KLOE. The respective statistical correlation matrices are compared in Fig.~\ref{fig:BaBar_stat} and Fig.~\ref{fig:KLOE_stat}. Both measurements include statistical correlations arising from their respective unfolding procedures, but also have other distinct sources of statistical correlation. In the case of KLOE, the KLOE08~\cite{KLOE:2008fmq} and KLOE12~\cite{KLOE:2012anl} measurements are statistically correlated as they utilize the same $\pi^+\pi^-$ data. However, the correlations are relatively weak as the final cross section in each case is derived from different approaches~\cite{KLOE-2:2017fda}. 

Statistical correlations in the BaBar measurement also arise from the use of the measured $e^+e^-\rightarrow\mu^+\mu^-(\gamma)$ luminosity technique~\cite{BaBar:2012bdw}. In this approach, the $\mu^+\mu^-$ process is used to normalize the measurement and derive $\sigma_{\pi\pi}$, benefiting from the cancellation of several systematic effects. This technique is widely used in radiative return (initial-state radiation based) measurements of $\sigma_{\pi\pi}$~\cite{BESIII:2015equ,Xiao:2017dqv, KLOE:2012anl, KLOE-2:2017fda}, including KLOE12~\cite{KLOE:2012anl}. However, while the BaBar $\pi^+\pi^-$ data are measured in fine 2~MeV bins over the $\rho$ resonance region ($0.5~\mathrm{GeV} < \sqrt{s} < 1.0~\mathrm{GeV}$), the corresponding $\mu^+\mu^-$ data used for normalization are measured in much coarser 50~MeV bins. While local variations are applied within these wider bins to better capture rapidly changing resonance features, such as $\rho$–$\omega$ interference, the use of widely binned $\mu^+\mu^-$ data induces strong statistical correlations between $\sigma_{\pi\pi}$ bins that share the same underlying normalization.

Although the KLOE measurements employ a more coarsely binned $\sigma_{\pi\pi}$ spectrum (10~MeV) in the $\rho$ resonance region compared to BaBar, there is a one-to-one correspondence between the $\pi\pi$ and $\mu\mu$ bins in the KLOE12 analysis. In addition, the KLOE10 measurement is statistically independent~\cite{KLOE:2010qei}, such that statistical correlations between KLOE datasets are relatively weak overall. The statistical correlations in the BaBar data are thus both stronger and longer-ranged than those in KLOE. Combined with its high statistical precision, BaBar’s statistical correlations are consequently a powerful experimental constraint on the $\pi^+\pi^-$ data combination. Reducing or eliminating these medium-range correlations in a data combination effectively neglects the underlying statistical structure, allowing strongly coupled bins to behave more independently in the fit. This, in turn, leads to the significant changes observed in the resulting spectrum.

In summary, while systematic correlations account for a small part of the difference between $a_\mu^{\pi^+\pi^-}$ as obtained by DHMZ19 and KNT19, it is evident that statistical correlations play the dominant role. 
In particular, the inclusion of the medium-range statistical correlations in the BaBar data reduces its effective weight in the combination and leads to a lower combined value, contributing to the difference historically attributed to KLOE (and other lower-lying datasets such as BESIII~\cite{BESIII:2015equ}). In the absence of correlations, BaBar’s statistical precision is such that it dominates the KNT19 fit in the $\pi\pi$ channel, as illustrated in Fig.~\ref{fig:both_decorr_spectra}. However, as emphasized above, statistical correlations are not subject to modeling assumptions in the same way as systematic correlations, but are instead derived directly from the underlying statistics of the measurements. As such, they should not be modified or reduced, even for the purpose of estimating uncertainties~\cite{Aoyama:2020ynm}. Tests show that the DHMZ result is recovered when BaBar’s medium-range, strongly constraining statistical correlations are diminished in conjunction with reduced systematic correlations. Such a procedure effectively neglects essential experimental information that governs the combined result for $a_\mu^{\pi^+\pi^-}$.

\bibliographystyle{apsrev4-1.bst}
\bibliography{bib/1_intro,bib/2_motivation,bib/3_procedure,bib/wp25}

\end{document}